\documentclass[aps,prd,twocolumn,amsmath,amssymb,bm,superscriptaddress,longbibliography,doi,epstopdf,numerical,preprintnumbers]{revtex4-1}  

\usepackage{calrsfs}
\usepackage{dsfont}
\DeclareMathAlphabet{\pazocal}{OMS}{zplm}{m}{n}

\newcommand{\Dsm}{\ensuremath{D_s^-}}

\newcommand{\Ds}{\ensuremath{D_s^\mp}}

\newcommand{\Dsdecay}{\ensuremath{D_s^\pm \rightarrow \phi \pi^{\pm}}}

\newcommand{\DsMdecay}{\ensuremath{D_s^- \rightarrow \phi \pi^{-}}}

\newcommand{\Bsdecay}{\ensuremath{B_s^0 \rightarrow \mu^+ D_s^- \, \mathrm{X}}} 
\newcommand{\Bspmdecay}{\ensuremath{B_s^0 \rightarrow \mu^\pm D_s^\mp \, \mathrm{X}}} 
\newcommand{\Bsjp}{\ensuremath{B_s^0 \rightarrow J/\psi\phi}}

\newcommand {\Bs} {\ensuremath{B^0_s}}

\newcommand {\sig} {\ensuremath{\sqrt{-2\, {\rm ln} ({\cal{L}}_0 /{\cal{L}}_\mathrm{max}) }}}

\usepackage{graphicx}
\usepackage{dcolumn}
\usepackage{color}
\usepackage{bm}
\usepackage[caption=false]{subfig}

\begin{document}

\preprint{FERMILAB-PUB-17-612-E}

\title{Study of the $\bm{X^\pm(5568)}$ state with  semileptonic decays of the $\bm{B_s^0}$ meson }

%
\affiliation{LAFEX, Centro Brasileiro de Pesquisas F\'{i}sicas, Rio de Janeiro, RJ 22290, Brazil}
\affiliation{Universidade do Estado do Rio de Janeiro, Rio de Janeiro, RJ 20550, Brazil}
\affiliation{Universidade Federal do ABC, Santo Andr\'e, SP 09210, Brazil}
\affiliation{University of Science and Technology of China, Hefei 230026, People's Republic of China}
\affiliation{Universidad de los Andes, Bogot\'a, 111711, Colombia}
\affiliation{Charles University, Faculty of Mathematics and Physics, Center for Particle Physics, 116 36 Prague 1, Czech Republic}
\affiliation{Czech Technical University in Prague, 116 36 Prague 6, Czech Republic}
\affiliation{Institute of Physics, Academy of Sciences of the Czech Republic, 182 21 Prague, Czech Republic}
\affiliation{Universidad San Francisco de Quito, Quito 170157, Ecuador}
\affiliation{LPC, Universit\'e Blaise Pascal, CNRS/IN2P3, Clermont, F-63178 Aubi\`ere Cedex, France}
\affiliation{LPSC, Universit\'e Joseph Fourier Grenoble 1, CNRS/IN2P3, Institut National Polytechnique de Grenoble, F-38026 Grenoble Cedex, France}
\affiliation{CPPM, Aix-Marseille Universit\'e, CNRS/IN2P3, F-13288 Marseille Cedex 09, France}
\affiliation{LAL, Univ. Paris-Sud, CNRS/IN2P3, Universit\'e Paris-Saclay, F-91898 Orsay Cedex, France}
\affiliation{LPNHE, Universit\'es Paris VI and VII, CNRS/IN2P3, F-75005 Paris, France}
\affiliation{CEA Saclay, Irfu, SPP, F-91191 Gif-Sur-Yvette Cedex, France}
\affiliation{IPHC, Universit\'e de Strasbourg, CNRS/IN2P3, F-67037 Strasbourg, France}
\affiliation{IPNL, Universit\'e Lyon 1, CNRS/IN2P3, F-69622 Villeurbanne Cedex, France and Universit\'e de Lyon, F-69361 Lyon CEDEX 07, France}
\affiliation{III. Physikalisches Institut A, RWTH Aachen University, 52056 Aachen, Germany}
\affiliation{Physikalisches Institut, Universit\"at Freiburg, 79085 Freiburg, Germany}
\affiliation{II. Physikalisches Institut, Georg-August-Universit\"at G\"ottingen, 37073 G\"ottingen, Germany}
\affiliation{Institut f\"ur Physik, Universit\"at Mainz, 55099 Mainz, Germany}
\affiliation{Ludwig-Maximilians-Universit\"at M\"unchen, 80539 M\"unchen, Germany}
\affiliation{Panjab University, Chandigarh 160014, India}
\affiliation{Delhi University, Delhi-110 007, India}
\affiliation{Tata Institute of Fundamental Research, Mumbai-400 005, India}
\affiliation{University College Dublin, Dublin 4, Ireland}
\affiliation{Korea Detector Laboratory, Korea University, Seoul, 02841, Korea}
\affiliation{CINVESTAV, Mexico City 07360, Mexico}
\affiliation{Nikhef, Science Park, 1098 XG Amsterdam, the Netherlands}
\affiliation{Radboud University Nijmegen, 6525 AJ Nijmegen, the Netherlands}
\affiliation{Joint Institute for Nuclear Research, Dubna 141980, Russia}
\affiliation{Institute for Theoretical and Experimental Physics, Moscow 117259, Russia}
\affiliation{Moscow State University, Moscow 119991, Russia}
\affiliation{Institute for High Energy Physics, Protvino, Moscow region 142281, Russia}
\affiliation{Petersburg Nuclear Physics Institute, St. Petersburg 188300, Russia}
\affiliation{Instituci\'{o} Catalana de Recerca i Estudis Avan\c{c}ats (ICREA) and Institut de F\'{i}sica d'Altes Energies (IFAE), 08193 Bellaterra (Barcelona), Spain}
\affiliation{Uppsala University, 751 05 Uppsala, Sweden}
\affiliation{Taras Shevchenko National University of Kyiv, Kiev, 01601, Ukraine}
\affiliation{Lancaster University, Lancaster LA1 4YB, United Kingdom}
\affiliation{Imperial College London, London SW7 2AZ, United Kingdom}
\affiliation{The University of Manchester, Manchester M13 9PL, United Kingdom}
\affiliation{University of Arizona, Tucson, Arizona 85721, USA}
\affiliation{University of California Riverside, Riverside, California 92521, USA}
\affiliation{Florida State University, Tallahassee, Florida 32306, USA}
\affiliation{Fermi National Accelerator Laboratory, Batavia, Illinois 60510, USA}
\affiliation{University of Illinois at Chicago, Chicago, Illinois 60607, USA}
\affiliation{Northern Illinois University, DeKalb, Illinois 60115, USA}
\affiliation{Northwestern University, Evanston, Illinois 60208, USA}
\affiliation{Indiana University, Bloomington, Indiana 47405, USA}
\affiliation{Purdue University Calumet, Hammond, Indiana 46323, USA}
\affiliation{University of Notre Dame, Notre Dame, Indiana 46556, USA}
\affiliation{Iowa State University, Ames, Iowa 50011, USA}
\affiliation{University of Kansas, Lawrence, Kansas 66045, USA}
\affiliation{Louisiana Tech University, Ruston, Louisiana 71272, USA}
\affiliation{Northeastern University, Boston, Massachusetts 02115, USA}
\affiliation{University of Michigan, Ann Arbor, Michigan 48109, USA}
\affiliation{Michigan State University, East Lansing, Michigan 48824, USA}
\affiliation{University of Mississippi, University, Mississippi 38677, USA}
\affiliation{University of Nebraska, Lincoln, Nebraska 68588, USA}
\affiliation{Rutgers University, Piscataway, New Jersey 08855, USA}
\affiliation{Princeton University, Princeton, New Jersey 08544, USA}
\affiliation{State University of New York, Buffalo, New York 14260, USA}
\affiliation{University of Rochester, Rochester, New York 14627, USA}
\affiliation{State University of New York, Stony Brook, New York 11794, USA}
\affiliation{Brookhaven National Laboratory, Upton, New York 11973, USA}
\affiliation{Langston University, Langston, Oklahoma 73050, USA}
\affiliation{University of Oklahoma, Norman, Oklahoma 73019, USA}
\affiliation{Oklahoma State University, Stillwater, Oklahoma 74078, USA}
\affiliation{Oregon State University, Corvallis, Oregon 97331, USA}
\affiliation{Brown University, Providence, Rhode Island 02912, USA}
\affiliation{University of Texas, Arlington, Texas 76019, USA}
\affiliation{Southern Methodist University, Dallas, Texas 75275, USA}
\affiliation{Rice University, Houston, Texas 77005, USA}
\affiliation{University of Virginia, Charlottesville, Virginia 22904, USA}
\affiliation{University of Washington, Seattle, Washington 98195, USA}
\author{V.M.~Abazov} \affiliation{Joint Institute for Nuclear Research, Dubna 141980, Russia}
\author{B.~Abbott} \affiliation{University of Oklahoma, Norman, Oklahoma 73019, USA}
\author{B.S.~Acharya} \affiliation{Tata Institute of Fundamental Research, Mumbai-400 005, India}
\author{M.~Adams} \affiliation{University of Illinois at Chicago, Chicago, Illinois 60607, USA}
\author{T.~Adams} \affiliation{Florida State University, Tallahassee, Florida 32306, USA}
\author{J.P.~Agnew} \affiliation{The University of Manchester, Manchester M13 9PL, United Kingdom}
\author{G.D.~Alexeev} \affiliation{Joint Institute for Nuclear Research, Dubna 141980, Russia}
\author{G.~Alkhazov} \affiliation{Petersburg Nuclear Physics Institute, St. Petersburg 188300, Russia}
\author{A.~Alton$^{a}$} \affiliation{University of Michigan, Ann Arbor, Michigan 48109, USA}
\author{A.~Askew} \affiliation{Florida State University, Tallahassee, Florida 32306, USA}
% ...
%
%
\collaboration{The D0 Collaboration}\thanks{with visitors from
$^{a}$Augustana College, Sioux Falls, SD 57197, USA,
$^{b}$The University of Liverpool, Liverpool L69 3BX, UK,
$^{c}$Deutshes Elektronen-Synchrotron (DESY), Notkestrasse 85, Germany,
$^{d}$CONACyT, M-03940 Mexico City, Mexico,
$^{e}$SLAC, Menlo Park, CA 94025, USA,
$^{f}$University College London, London WC1E 6BT, UK,
$^{g}$Centro de Investigacion en Computacion - IPN, CP 07738 Mexico City, Mexico,
$^{h}$Universidade Estadual Paulista, S\~ao Paulo, SP 01140, Brazil,
$^{i}$Karlsruher Institut f\"ur Technologie (KIT) - Steinbuch Centre for Computing (SCC),
D-76128 Karlsruhe, Germany,
$^{j}$Office of Science, U.S. Department of Energy, Washington, D.C. 20585, USA,
$^{k}$American Association for the Advancement of Science, Washington, D.C. 20005, USA,
$^{l}$Kiev Institute for Nuclear Research (KINR), Kyiv 03680, Ukraine,
$^{m}$University of Maryland, College Park, MD 20742, USA,
$^{n}$European Orgnaization for Nuclear Research (CERN), CH-1211 Geneva, Switzerland,
$^{o}$Purdue University, West Lafayette, IN 47907, USA,
$^{p}$Institute of Physics, Belgrade, Belgrade, Serbia,
and
$^{q}$P.N. Lebedev Physical Institute of the Russian Academy of Sciences, 119991, Moscow, Russia.
$^{\ddag}$Deceased.
} \noaffiliation
\vskip 0.25cm

\date{28 December 2017}

\begin{abstract}
We present a study of the $X^\pm(5568)$ using semileptonic decays of the
$B_s^0$ meson using the full run II integrated luminosity of 10.4
fb$^{-1}$ in proton-antiproton collisions at a center of mass energy of 1.96\,TeV 
collected with the D0 detector
at the Fermilab Tevatron Collider. We report evidence for a narrow
structure, $X^\pm(5568)$, in the decay sequence $X^\pm(5568) \to B_s^0 \pi^\pm$
where $B_s^0 \rightarrow \mu^\mp D_s^\pm \, \mathrm{X}$,
\Dsdecay\ which is consistent with the previous measurement  by  the D0
Collaboration in the hadronic  decay mode, $X^\pm(5568) \to \Bs\pi^\pm$  where
$\Bs \to J/\psi\phi$. The mass and   width of this
state are measured using a combined fit of the hadronic and semileptonic data, 
yielding $m =
5566.9 ^{+3.2}_{-3.1} \thinspace  {\rm (stat)} ^{+0.6}_{-1.2}  {\rm
\thinspace (syst)}$\,MeV/$c^2$, $\Gamma = 18.6 ^{+7.9}_{-6.1}  {\rm
\thinspace (stat)}   ^{+3.5}_{-3.8} {\rm \thinspace (syst)} $\,MeV/$c^2$
with a  significance of 6.7$\,\sigma$.
\end{abstract}

\maketitle

\section{Introduction}

Since the creation of the quark
model~\cite{GellMann:1964nj,Zweig:1981pd} it was understood that exotic
mesons containing more than one quark-antiquark pair are possible.
However, for exotic mesons containing only the up, down and strange
quarks it has been difficult to make a definitive experimental case for
such exotic states, although some persuasive arguments have been
made
(for recent comprehensive discussions of exotic hadrons 
containing both light and heavy quarks, see Refs.~\cite{Maiani:2004uc, Esposito:2016noz, Chen:2016qju, 
[{}][{}]Olsen:2017bmm}).
Multiquark states
that contain heavy quarks can be more recognizable owing to the
distinctive decay structure of heavy quark hadrons. The 2003 discovery
by the Belle experiment~\cite{Choi:2003ue} of the $X(3872)$ in the
channel $B^\pm \to K^\pm X (\to \pi^+ \pi^- J/\psi)$ was the first
candidate exotic meson in which heavy flavor quarks participate. This
state was subsequently confirmed in several production and decay modes
by ATLAS~\cite{Aaboud:2016vzw}, BaBar~\cite{Aubert:2004ns}, BES
III~\cite{Ablikim:2013dyn}, CDF~\cite{Acosta:2003zx},
CMS~\cite{Chatrchyan:2013cld}, D0~\cite{Abazov:2004kp} and
LHCb~\cite{Aaij:2013zoa} Collaborations. Several additional four-quark candidate exotic
mesons have since been found, though in many cases not all experiments
have been able to confirm their existence.

Four-quark mesons can be generically categorized as either ``molecular states''
or tetraquark  states of a diquark and an anti-diquark. In the example
of the $X(3872)$, a molecular state interpretation would be a colorless
$D^0$ ($c \bar u$) and a colorless $\bar{D}^{*0}$ ($u \bar c$) in a
loosely bound state. Such a state would be expected to lie close in mass
to the $D^0 \bar{D}^{*0}$ threshold. The tetraquark mode of a colored
diquark ($c u$) and colored anti-diquark ($\bar c \bar u$) is more
strongly bound by the exchange of  gluons and would be expected to have
a mass somewhat below the $D^0$$\bar{D}^{*0}$ threshold. In many cases,
interpretations of four-quark mesons as pure molecular or tetraquark
states are difficult and more complex mechanisms may be
required~\cite{Esposito:2016noz, Chen:2016qju,Olsen:2017bmm}. 
The firm
identification of multiquark mesons and baryons and the study of their
properties are of importance for further understanding of
nonperturbative QCD.

Recently the D0 Collaboration presented evidence for a new four-quark
candidate that decays to  $\Bs\pi^\pm$ where \Bs\ decays to $J/\psi \phi$~\cite{D0:2016mwd}.
This system would be composed of two quarks and two antiquarks of four
different flavors: $b, s, u, d$, with either a molecular constitution as a
loosely bound $B_d^0$ and $K^\pm$ system or a tightly bound tetraquark
such as  $(bd)$-$(\bar s \bar u)$,  $(bu)$-$(\bar s \bar d)$, 
$(su)$-$(\bar b \bar d)$, or  $(sd)$-$(\bar b \bar u)$ (because the \Bs\
meson is fully mixed, the exact quark anti-quark composition cannot be
determined). The mass of $X^\pm(5568)$ is about 200\,MeV/$c^2$ below the
$B_d^0 K^\pm$ threshold, thus disfavoring a $B_d^0$-$K^\pm$ molecular
interpretation.

The $X^\pm(5568)$ was previously reported~\cite{D0:2016mwd} with a significance 
of 5.1\,$\sigma$ (including systematic uncertainties and the look-elsewhere 
effect~\cite{[{For a description  of the  look-elsewhere 
effect see }][{}]lyons2008}) in the decay $X^\pm(5568) \to \Bs (J/\psi \phi) \pi^{\pm}$
 in proton-antiproton collisions at a center of mass energy of 1.96\,TeV. 
The ratio of the number of $B_s^0$ that are from the decay of the $X^\pm(5568)$ 
to all $B_s^0$ produced was  measured to be $\left[8.6 \pm 1.9 \thinspace  {\rm (stat)}
\pm 1.4 \thinspace  {\rm (syst)}\right]\%$. 
In order to reduce the background, a selection was imposed on
the angle between the 
$B_s^0$ 
and $\pi^{\pm}$ (the ``cone cut'', $\Delta R = \sqrt{\Delta\eta^2 +
\Delta\phi^2} <0.3$~\footnote{$\eta=-\ln[\tan(\theta/2)]\ $ is the
pseudorapidity and $\theta$ is the polar angle between the track
momentum and the proton beam direction. $\phi$ is the azimuthal angle of
the track.}).
Without the cone cut the significance was found to be 3.9\,$\sigma$. 
In addition to increasing the signal-to-background ratio this cone cut
limits  backgrounds, 
such as possible excited states of the $B_c$ meson, that
are not included in the available simulations. Multiple checks were carried out to
ensure that the cone cut did not create an anomalous signal~\cite{[{}] [{ (and the
Supplemental Material at \protect
{\tt{\url{https://journals.aps.org/prl/supplemental/10.1103/PhysRevLett.117.022003}}}
for additional figures) }]D0:2016mwd}. Varying the  cone cut from
$\Delta R_\mathrm{max} = 0.2$ to $0.5$ gave stable fitted masses and resulted in  no unexpected changes in the
result. The invariant mass spectra of the $B_s^0$ candidates and charged
tracks with kaon or proton mass hypotheses were checked, and no resonant
enhancements in these distributions were found. The invariant mass
distribution of $B_d^0 \pi^\pm$ was also examined with no unexpected
resonances or reflections found. 
Subsequent analyses by the LHCb Collaboration~\cite{Aaij:2016iev} and 
by the CMS Collaboration~\cite{[{}][{ Phys. Rev. Lett. (to be published)}]Sirunyan:2017ofq} have not
found evidence for the  $X^\pm(5568)$ in proton-proton interactions at $\sqrt{s} = 7$
and 8\,TeV.  
The CDF Collaboration  has recently reported no evidence for $X^\pm (5568)$ in proton-antiproton collisions at $\sqrt{s} = 1.96$\,TeV~\cite{[{}][{ submitted to Phys. Rev. Lett.}]Aaltonen:2017voc} with different kinematic coverage than that of Ref.~\cite{D0:2016mwd}.

\begin{figure}[htbp]
  \includegraphics[width=\linewidth]{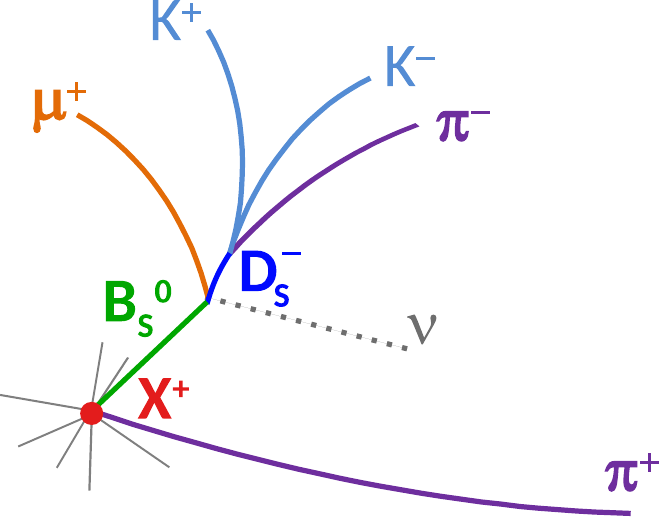}
\caption{\label{fig:decay} 
An illustration of the decay $X^+(5568) \to \Bs\pi^+$ where $\Bsdecay$ 
in the plane perpendicular to the beam.
}
\end{figure}

In this article, we present a study of the  $X^\pm(5568)$ in the decay to
$\Bs\pi^\pm$ using semileptonic \Bs\ decays,  \Bsdecay,  where \DsMdecay,
$\phi \to K^+K^-$, using the full run II integrated luminosity of 10.4
fb$^{-1}$ in proton-antiproton collisions at a center of mass energy of 1.96\,TeV 
collected with the D0 detector
at the Fermilab Tevatron Collider. Charge conjugate states are assumed. 
Here X  includes the unseen neutrino and
possibly a photon or $\pi^0$ from a $D_s^\ast$ decay or other hadrons
from the \Bs\ decay. The decay process is illustrated in
Fig.~\ref{fig:decay}. The semileptonic  decay channel has a higher
branching fraction than  the hadronic channel (\Bsjp). However the
presence of the unmeasured neutrino in the final state  deteriorates the
mass resolution of the signal. Still, a good mass resolution for the $X^\pm(5568)$
can  be obtained  in the semileptonic channel for events with a  large
invariant  mass of the $\mu^+ D_s^-$ system, yielding a comparable number 
of selected \Bs\ candidates in the two channels.
The backgrounds in the semileptonic channel are
independent of, but somewhat larger than, those in the hadronic channel. 
The character of possible reflections
of other resonant structures is quite different in the semileptonic and
hadronic channels. Thus observation  of the $X^\pm(5568)$ in the semileptonic decay
channel enables an independent confirmation of its existence. We report
here the results of the search for the $X^\pm(5568)$ in the semileptonic channel, as well as a
combination of the results in the hadronic and semileptonic channels. 

\section{D0 detector}

The detector components most relevant to this analysis are the central
tracking and the muon systems. The D0 detector has a central tracking
system consisting of a silicon microstrip tracker (SMT) and a central
fiber tracker (CFT), both located within a 2\,T superconducting
solenoidal magnet~\cite{Abazov:2005pn, Angstadt:2009ie}. The SMT has a
design optimized for tracking and vertexing for pseudorapidity of
$|\eta|<3$. For charged particles, the resolution on the distance of
closest approach as provided by the tracking system is approximately
50\,$\mu$m for tracks with $p_T \approx 1$~GeV/$c$, where $p_T$ is the
component of the momentum perpendicular to the beam axis. It improves
asymptotically to 15~$\mu$m for tracks with $p_T > 10$~GeV/$c$.
Preshower detectors and electromagnetic and hadronic calorimeters
surround the tracker. A muon system, positioned outside the calorimeter,
covering $|\eta|<2$ consists of a layer of tracking detectors and
scintillation trigger counters in front of 1.8\,T iron toroidal magnets,
followed by two similar layers after the toroids~\cite{Abazov:2005uk}.

\section{Event reconstruction and selection} 
 
The \Bsdecay\  selection requirements have been chosen to optimize
the mass resolution of the $\Bs\pi^+$ system and to minimize background
from random combinations of tracks from muons and  charged hadrons.  
The selection criteria are based on those used in Ref.~\cite{Abazov:2012zz}
with the cut on the \Bs\ isolation removed and have been selected by maximizing 
the significance of the signal.

The data were collected with a suite of single and dimuon triggers
(approximately 95\% of the sample is recorded using single muon
triggers). The selection and reconstruction of $\mu^{+} D_s^{-}$ decays
requires tracks with at least two hits in both the CFT and SMT.


The muon is required to have hits in at least two layers of the muon
system, with   segments reconstructed both  inside and outside the
toroid. The muon track segment is required to be matched to a track
found in the central tracking system that has transverse momentum $3 <
p_T < 25$\,GeV/$c$. 

The $D_s^- \rightarrow  \phi \pi^-$; $\phi \rightarrow K^+ K^-$ decay is
selected as follows. The two particles from the $\phi$ decay are
assumed to be kaons and are required to have  $p_T > 1.0$\,GeV/$c$,
opposite charge and an invariant mass $1.012 < m(K^+K^-) < 1.03$\,GeV/$c^2$. The
charge of the third particle, assumed to be a pion, has to be opposite
to that of the muon. This particle is required to have transverse momentum $0.5 < p_T <25$\,GeV/$c$.
The mass of the three particles must satisfy  
 $1.91 < m(K^+K^-\pi^-) < 2.03$\,MeV$/c^2$. The
three tracks are combined to form a common $D_s^-$ decay vertex using
the algorithm described in Ref.~\cite{Abdallah:2002xm}. The $D^-_s$
vertex is required to be displaced from the $p\bar{p}$ primary
interaction vertex (PV) in the transverse plane with a significance of
at least three standard deviations. The cosine of the angle between the
$D^-_s$ momentum and the vector from the PV to the $D^-_s$ decay vertex
is required to be greater than 0.9.

The trajectories of the muon and $D^-_s$ candidate are required to be
consistent with originating from a common vertex (assumed to be the
$B^0_s$ semileptonic decay vertex). The cosine of the angle between the
combined $\mu^+ D^-_s$ transverse momentum, an approximation of the
$B^0_s$ direction, and the direction from the PV to the $B^0_s$ decay
vertex has to be greater than 0.95. The \Bs\ decay vertex has to be
displaced from the PV in the transverse plane with a significance of at
least four standard deviations. The transverse momentum of the $\mu^+
D_s^-$ system is required to satisfy the condition $p_T > 10$\,GeV/$c$ 
to suppress backgrounds. To minimize the effect of the neutrino in the
final state the effective mass is limited to $4.5\,$GeV$/c^2\, < m(\mu^+
D_s^-) < m(\Bs)$. 

The impact parameters (IP)~\footnote{The three dimensional impact
parameter (IP) is defined as the distance of closest approach of the
track to the $p\bar{p}$ collision point. The two dimensional IP is the
distance of closest approach projected onto the plane transverse to the
$p\bar{p}$ beams.} with respect to the PV of the four tracks from the
\Bs\ decay are required to satisfy the following criteria: the
two-dimensional (2D) IPs of the tracks of the muon and the pion from  the
\Dsm\ decay are required to be at least 50\,$\mu$m  to reject tracks
emerging promptly from the PV (this requirement is not applied  to the tracks
associated with the charged kaons since the mass requirements provide
satisfactory background suppression). The three-dimensional (3D) IPs of
all four tracks are required to be less than 2\,cm to suppress
combinations with tracks emerging from different $p\bar{p}$ vertices
reconstructed in the same beam crossing.

\begin{figure}[htbp]
  \includegraphics[width=\columnwidth]{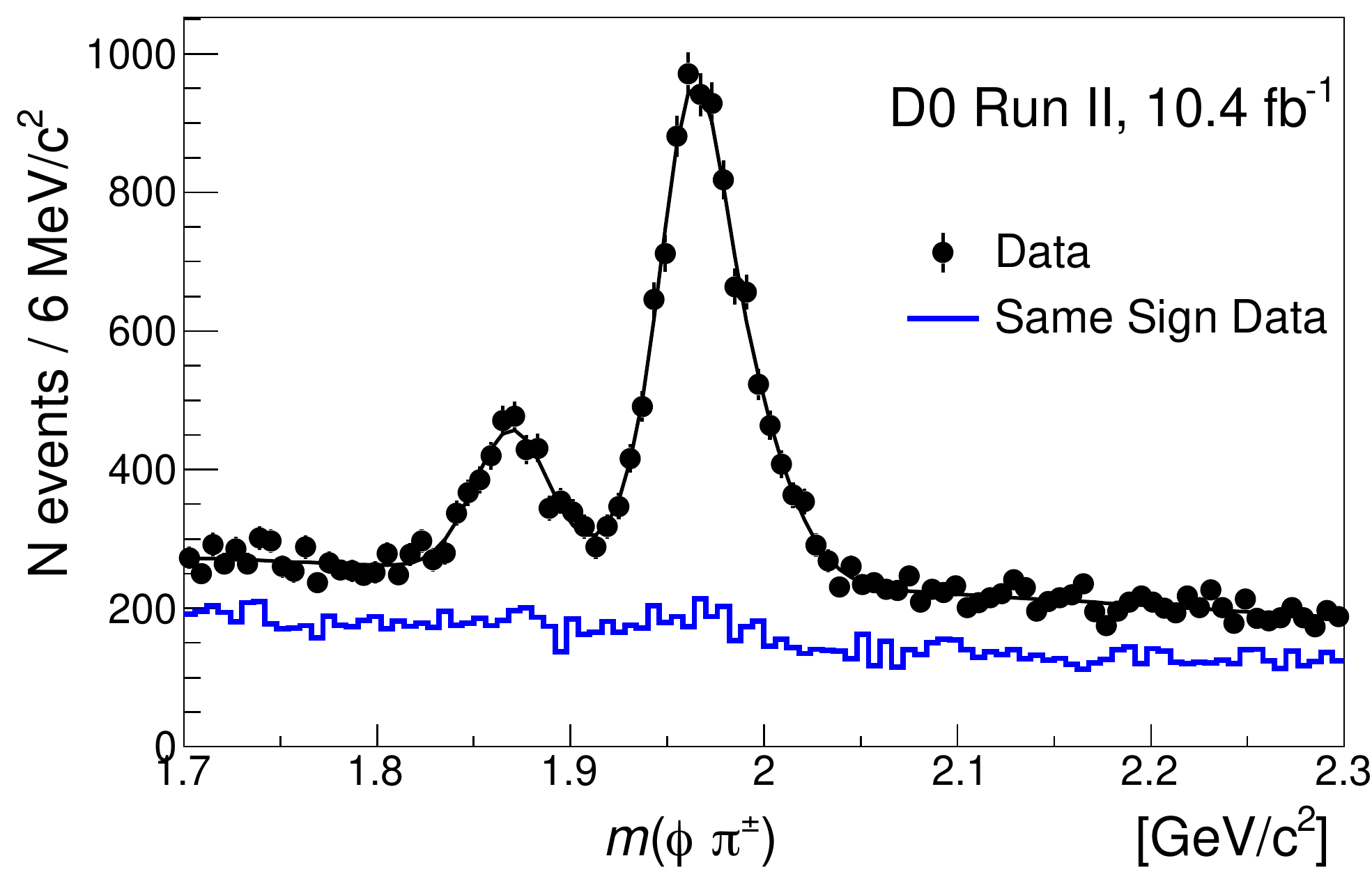}
\caption{\label{fig:ad3} 
The $K^+K^-\pi^\pm$ invariant mass distribution for
 the $\mu^{\pm}\phi\pi^\mp$  sample (right sign) with the  solid curve representing the fit. 
 The lower mass peak is due to  the decay $D^{\pm} \rightarrow \phi \pi^\pm$ and the second peak 
 is due to the $D_s^{\pm}$  decay. The blue histogram below the data points is the invariant mass distribution for
 the same-sign sample,  $\mu^{\pm}\phi\pi^\pm$.
}
\end{figure}

The $m(K^+K^-\pi^\pm)$  distribution of the candidates
that pass these cuts [except  $1.91 < m(K^+K^-\pi^-) < 2.03$\,MeV$/c^2$] 
is shown in Fig.~\ref{fig:ad3}, where   the
invariant mass distribution in data is compared to a fit using
a function which includes three terms: a second
order polynomial used to describe combinatorial background, a Gaussian
used to model the $D^{-}$ peak, and a double Gaussian with similar, but
different masses and  widths used to model the \Dsm\ peak. 

\begin{figure}[htbp]
  \includegraphics[width=\columnwidth]{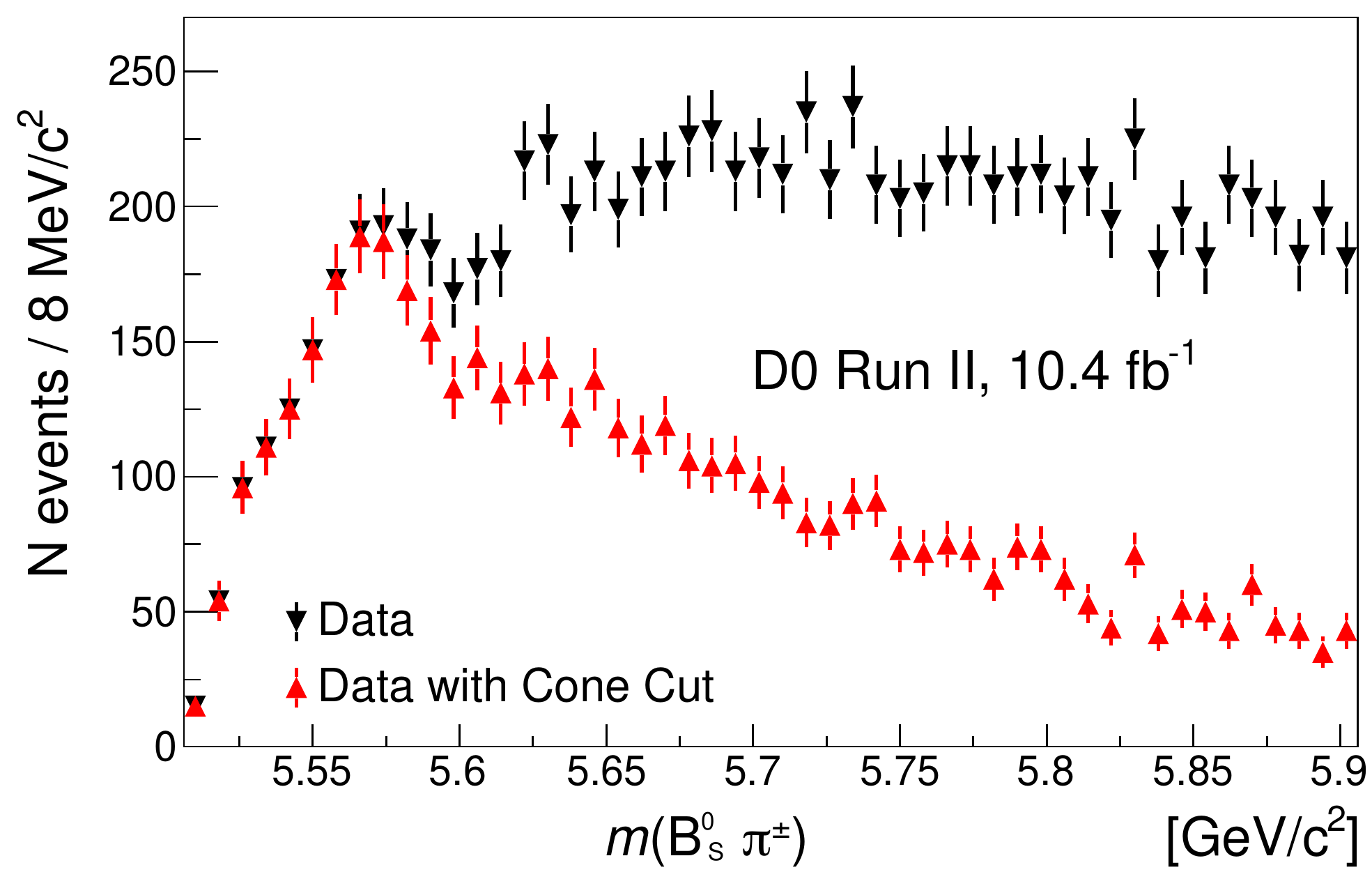}
\caption{\label{fig:massDist} 
The $m(\Bs\pi^\pm)$ distribution for the semileptonic data with (red upward triangles) and without (black downward triangles)  the cone cut. 
Below 5.56\,GeV/$c^2$ the red and black points have the same values. 
}
\end{figure}

The selection criteria for  the  pion in the $\Bs\pi^\pm$ combination 
have been chosen to match those used in the hadronic analysis. 
The track representing the pion 
is
required to have transverse momentum $0.5 < p_T <25$\,GeV/$c$ (the upper
limit is applied to reduce background). The pion and the \Bs\ candidate
are combined to form a  vertex that is consistent with the PV. The  pion
is required to be associated with the PV and have a 2D IP of at most
200\,$\mu$m and a 3D IP that is less than 0.12\,cm. Events with more
than 20 $\Bs\pi^\pm$ candidates are rejected. 
The most likely number of candidates per event is 5.1 and only about 0.1\% of the events have more than 20 candidates per event.
To improve the resolution
of the invariant mass of the $\Bs\pi^\pm$ system we define the invariant
mass as $m(\Bs\pi^\pm) = m(\mu^+\Dsm\pi^\pm) - m(\mu^+\Dsm) + m(\Bs) $
where $m(\Bs) = 5.3667$\,GeV/$c^2$~\cite{Olive:2016xmw}. We study 
the mass distribution in the range $5.506 < m(\Bs\pi^\pm) < 5.906$\,GeV/$c^2$. When using the hadronic data from
Ref.~\cite{D0:2016mwd} in this paper we use the same mass range as the 
semileptonic data instead of the slightly shifted mass range used in the original analysis
($5.5 < m(\Bs\pi^\pm) < 5.9$\,GeV/$c^2$). 
The semileptonic data are studied with and
without a cone cut which is used to suppress background, in which the angle between the $\mu^+ D_s^-$ system
and $\pi^{\pm}$ is required to satisfy $\Delta R = \sqrt{\Delta\eta^2 +
\Delta\phi^2} <0.3$.
The
resulting invariant mass distributions for the semileptonic channel are shown in
Fig.~\ref{fig:massDist}.

The selection cuts and resulting kinematics  for the hadronic and semileptonic  channels are quite similar.
The requirement that muons be seen outside the toroids means that the minimum $p_T$  
for the $J/\psi$ in the hadronic channel  is about 4\,GeV and about 3\, GeV for the
single muon in the semileptonic channel.
The minimum $p_T$ for the additional pion 
 is 0.5\,GeV for both the hadronic and semileptonic channels. 
For both channels, we require the minimum $p_T(\Bs\pi)$ to be greater than 10\,GeV and
the average $p_T(\Bs\pi)$ for events with  $m(B_s \pi) \approx 5.5$\,GeV is $\approx17$\, GeV.
For both channels the $\Bs\pi$ candidates are in the range of $-2 < \eta <2$ and more than
half of the events have a muon with  $|\eta|>1$.

\section{Monte Carlo simulation, Background Modeling and  parameterization}
\label{sec:background}

Monte Carlo (MC) samples are  generated using the {\sc
pythia}~\cite{Sjostrand:2006za}  event generator, modified to use {\sc
evtgen}~\cite{[{}] [{ for details see
http://www.slac.stanford.edu/$\sim$lange/EvtGen}]Lange:2001uf} for the
decay of hadrons containing $b$ or $c$ quarks.
The
generated events are processed by the full detector simulation chain.
Data events recorded in random
beam crossings are overlaid on the MC events to simulate the effect of
additional collisions in the same or nearby bunch crossings. 
The resulting events are then processed with the same reconstruction and selection algorithms as used
for data events.

\begin{figure*}[htb]
\includegraphics[width=0.49\linewidth]{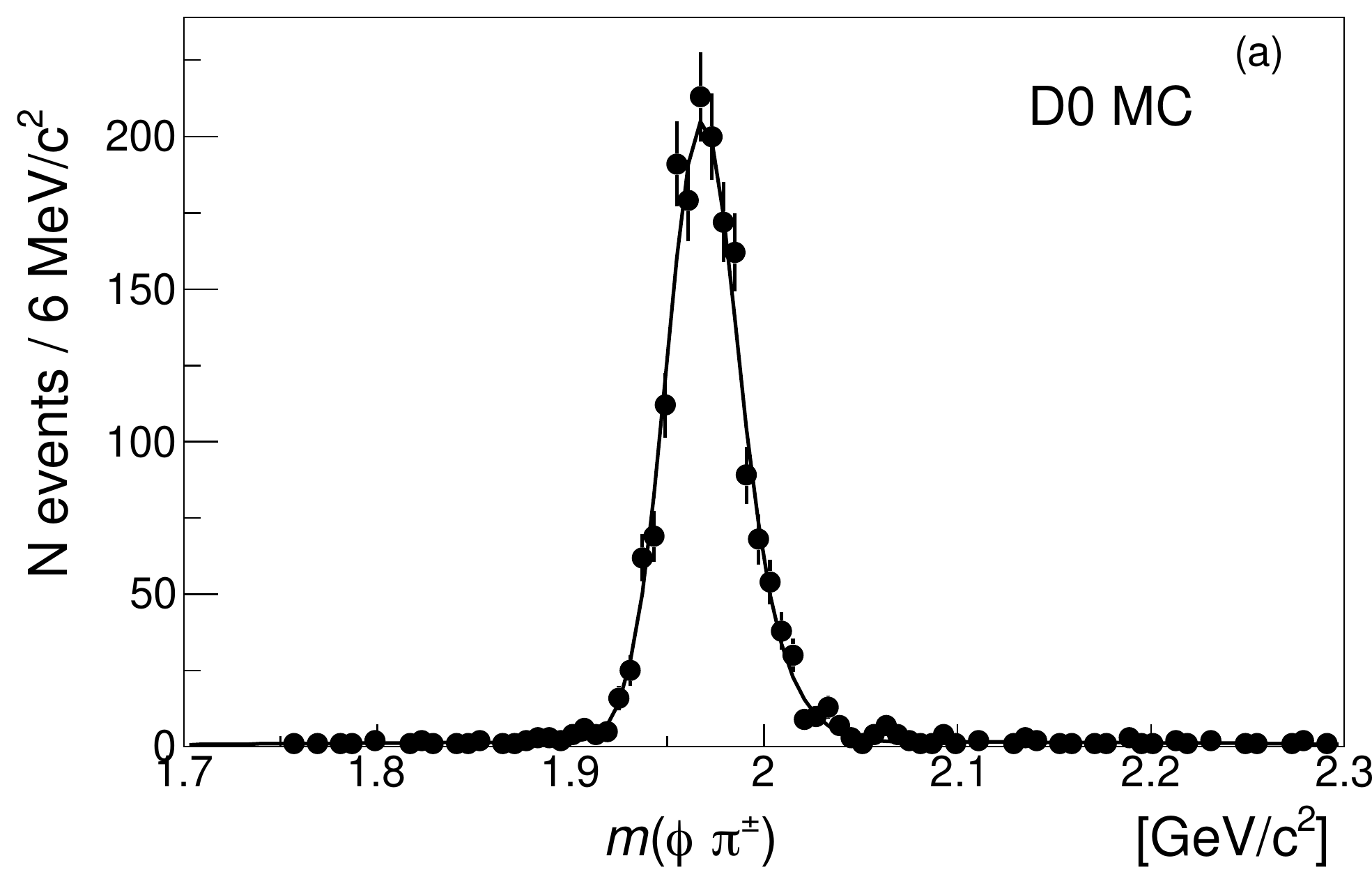}
\includegraphics[width=0.49\linewidth]{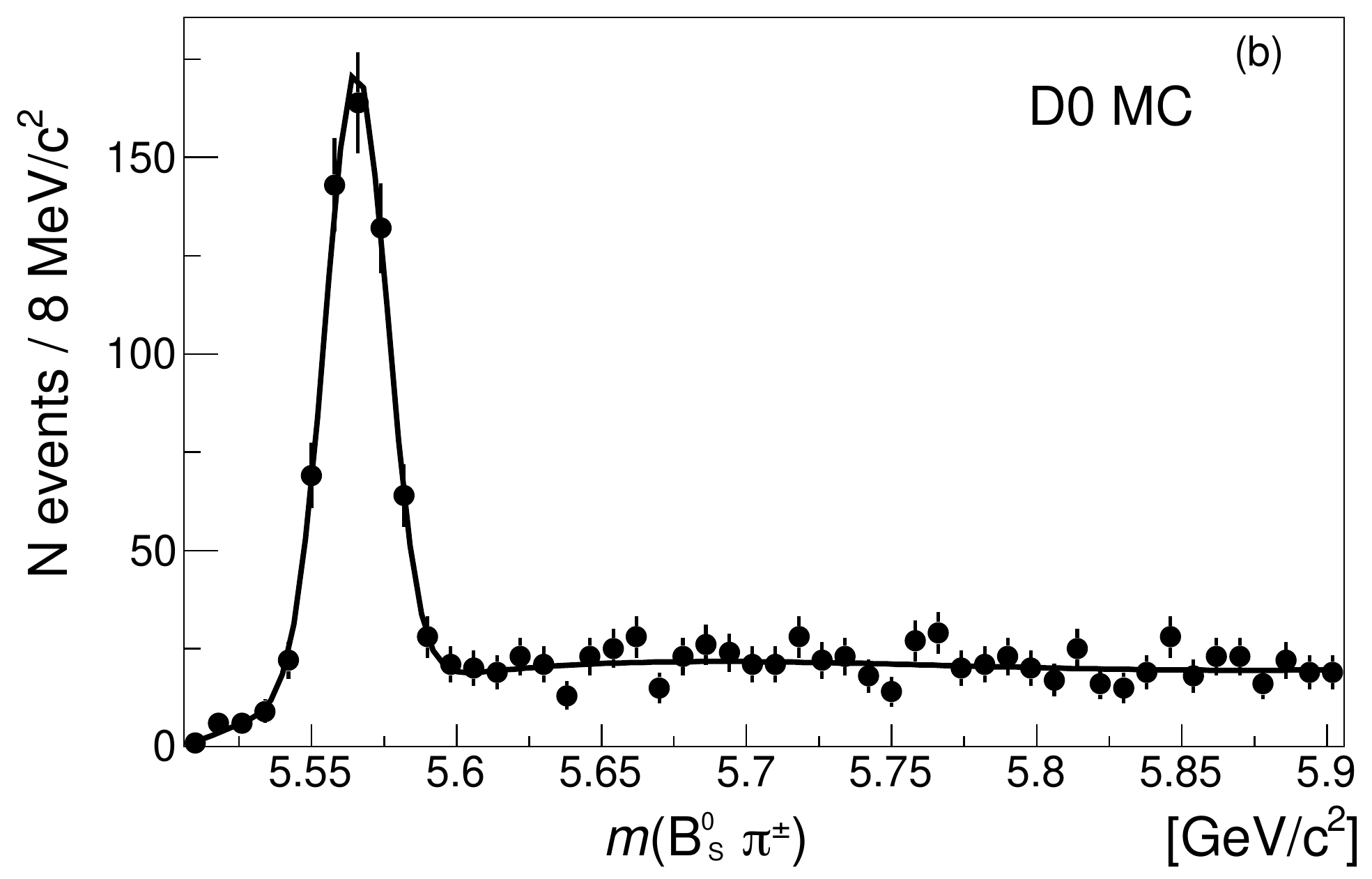}
\caption{MC simulation of $X^\pm(5568) \to \Bs\pi^\pm$ where \Bsdecay\ and the width of the 
$X^\pm(5568)$ is zero.
The invariant mass distributions  a) $m(K^+K^-\pi^-)$ and b) $m(\Bs\pi^+)$ are shown. 
The background in the  $m(\Bs\pi^+)$ distribution is  
produced by the combination of a random charged track with the \Bs\ meson.
\label{fig:signalMC}}
\end{figure*}

The MC sample for $X^\pm(5568)$ signal  is generated by modifying the mass
of the $B^\pm$ meson and forcing it to decay to $\Bs\pi^\pm$  using an
isotropic S-wave decay model. The $X^\pm(5568)$ is simulated with zero
width and zero lifetime. The resulting $K^+K^-\pi^-$  and $\Bs\pi^\pm$  invariant
mass distributions are shown in Fig.~\ref{fig:signalMC} with all selection requirements. 

The signal component of the $K^+K^-\pi^\pm$ invariant mass distribution
(Fig.~\ref{fig:signalMC}\,a) is modeled by two Gaussian functions and
the background by a second-order  polynomial. The signal of the
$m(\Bs\pi^\pm)$ distribution (Fig.~\ref{fig:signalMC}\,b) is well modeled
with a single Gaussian and the background with a third-order 
polynomial times an exponential. Using the results of these fits the
reconstruction efficiency of  the charged pion in the decay $X^\pm(5568) \to
\Bs\pi^\pm$ 
is $\left[ 32.0 \pm  1.8 \thinspace  {\rm (stat)} \pm 1.6 \thinspace  {\rm (syst)} \right]\%$
for $p_T (\mu^+\Dsm) > 10$\,GeV/$c$ where the systematic uncertainty
represents the expected differences between the reconstruction efficiencies for
low-momentum tracks in the  MC simulation and data.

It is not possible to create a model of the background that is based only on data. 
Since the $X^\pm(5568)$ decays to \Bs\ mesons, any data sample that includes \Bs\ decays 
will also include the signal and is unsuitable for modeling the background. 
Hence, we use MC-generated \Bs\ events that result from known particles that have
decays that include a \Bs\ in the decay chain, combined with data events where the muon has 
the same sign as 
the \Dsm\ candidate (SS events). MC event generators do not include all possible 
states as in many cases they have not been 
experimentally observed. For example,  
$b \bar{c}$ resonances decaying to \Bs\ mesons could contribute to our sample.

There are two distinct sources of background in this analysis. The 
first occurs when an  $X^\pm(5568)$ candidate is reconstructed from  
a real $\mu^+$ and $\Dsm$  together with a random charged track. This 
background is  modeled using  MC samples.

\begin{figure*}[hbtp]
\includegraphics[width=0.49\linewidth]{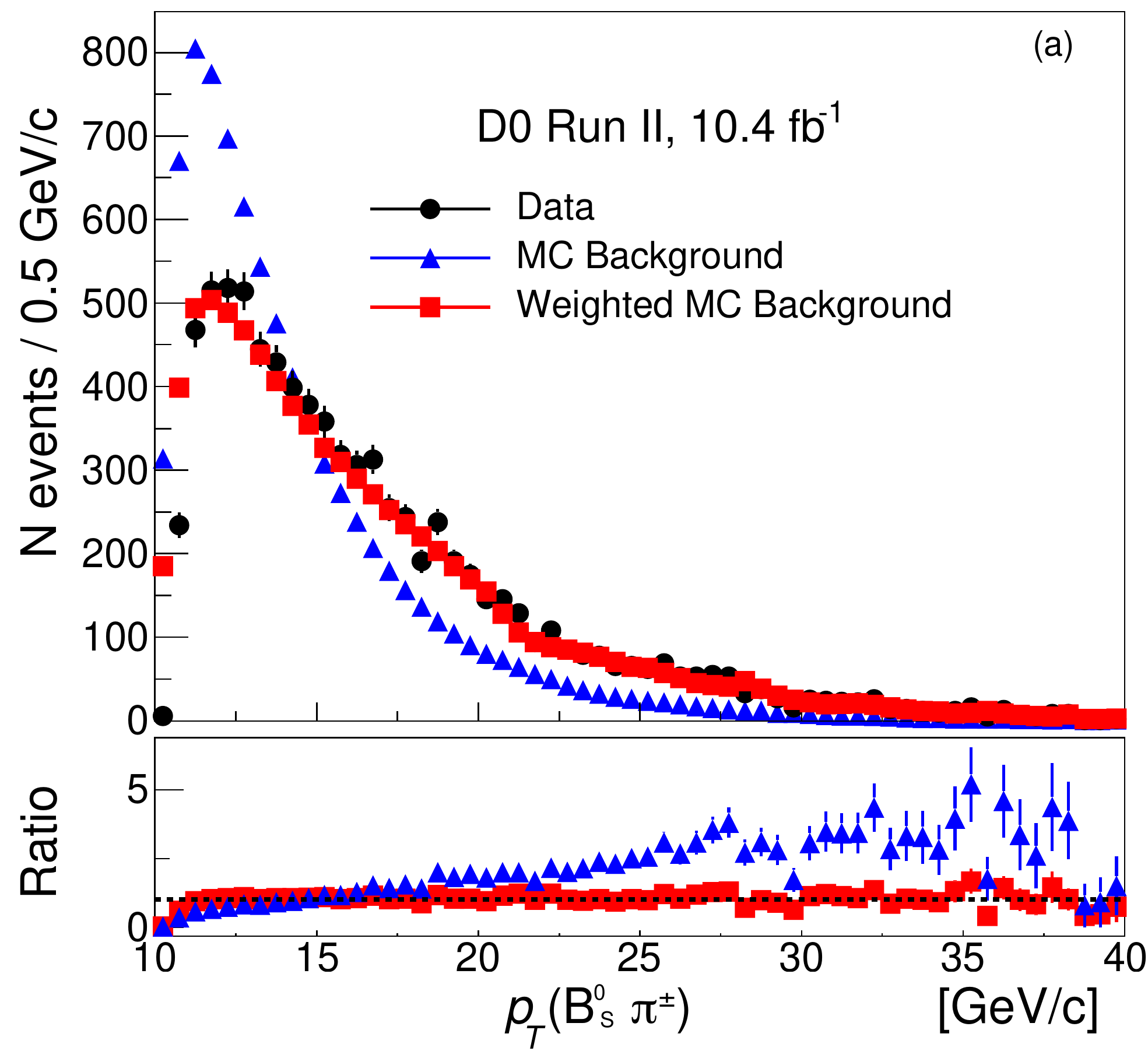}
\includegraphics[width=0.49\linewidth]{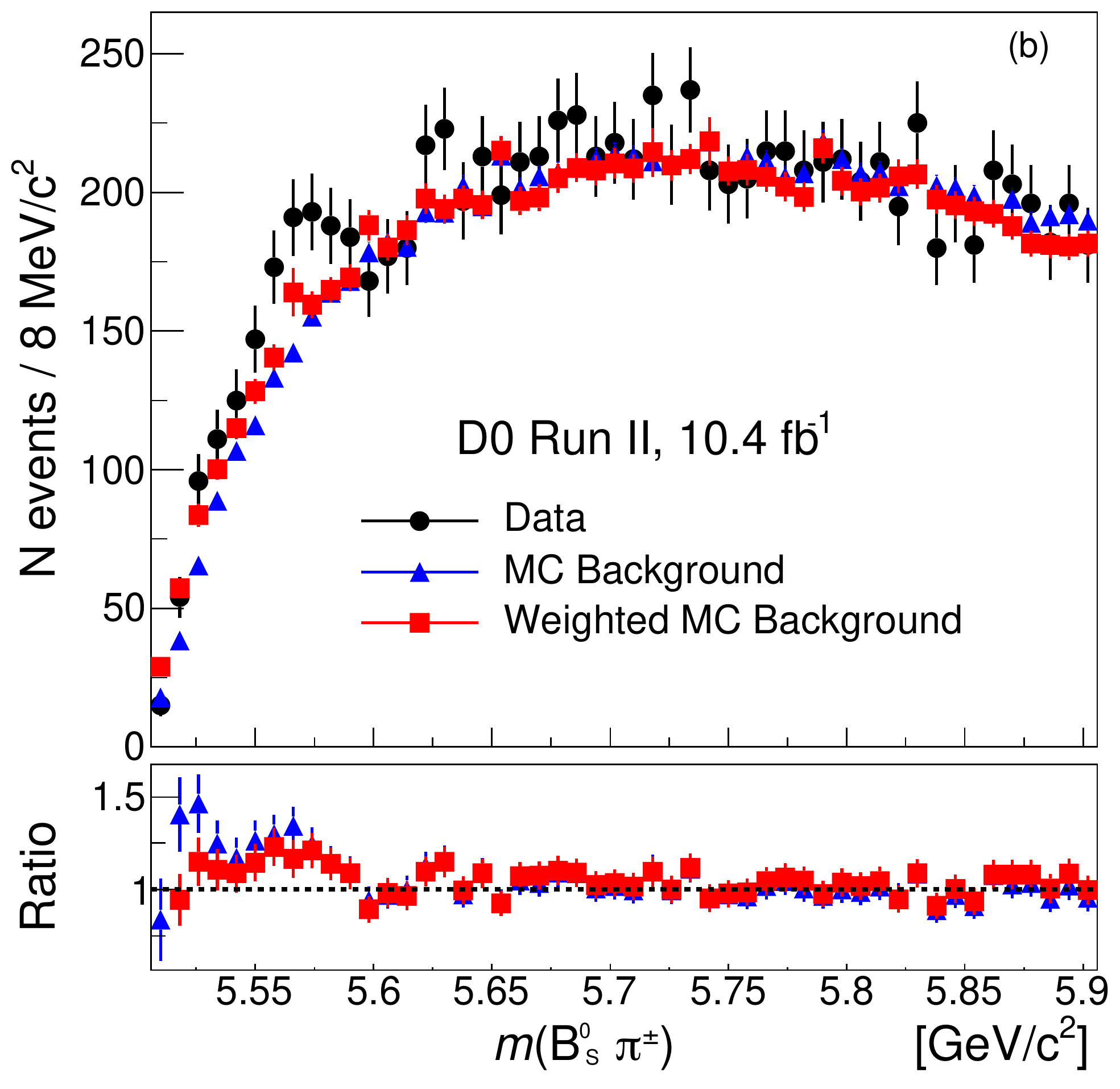}
\caption{
The MC background distribution, without the cone cut, before and after weighting is compared
with data (black points). The unweighted MC simulationis in blue and the weighted
is in red. The a) $p_T(\Bs\pi^\pm)$  and  b) invariant mass distributions
 $m(\Bs\pi^\pm)$ are shown. The excess in the data around
$m(\Bs\pi^\pm) = 5565$\,MeV/c is the $X^\pm(5568)$ signal.  The lower panels show the
ratio between the data and corresponding MC simulation.
\label{fig:weighted}}
\end{figure*}

The background MC sample is generated using the {\sc pythia}  inclusive
heavy flavor production model and events are selected that contain at least one
muon and a $D_s^\mp \rightarrow  \phi \pi^\mp$ decay where $\phi \rightarrow K^+ K^-$. 
To correct for the difference in lifetimes in the MC simulation and data, 
a weighting is applied to all nonprompt events in the
simulation, based on the generated lifetime of the $B$ candidate, to
give the world-average $B$ hadron lifetimes~\cite{Olive:2016xmw}. 
To correct for the effects of
the trigger selection and the reconstruction in data, we also weight each
MC event so that the transverse momenta of  the reconstructed muon  and
the $\mu^+\Dsm$ system agree with those in the data. The $p_T$
distribution of the $\Bs\pi^\pm$ system  is altered significantly by the weighting as shown in
Fig.~\ref{fig:weighted}(a). However, the effect is relatively small for the
$\Bs\pi^\pm$ mass distribution as seen in Fig.~\ref{fig:weighted}(b).

The second source of background is the combinatorial 
background that occurs when a $X^\pm(5568)$ candidate  is reconstructed from a 
spurious \Dsm\ candidate formed from three random charged tracks that form 
a vertex. This background is modeled using data events where the muon has the same sign as 
the \Dsm\ candidate (SS events). 

\begin{figure*}[htbp]
  {\includegraphics[width=0.49\linewidth]{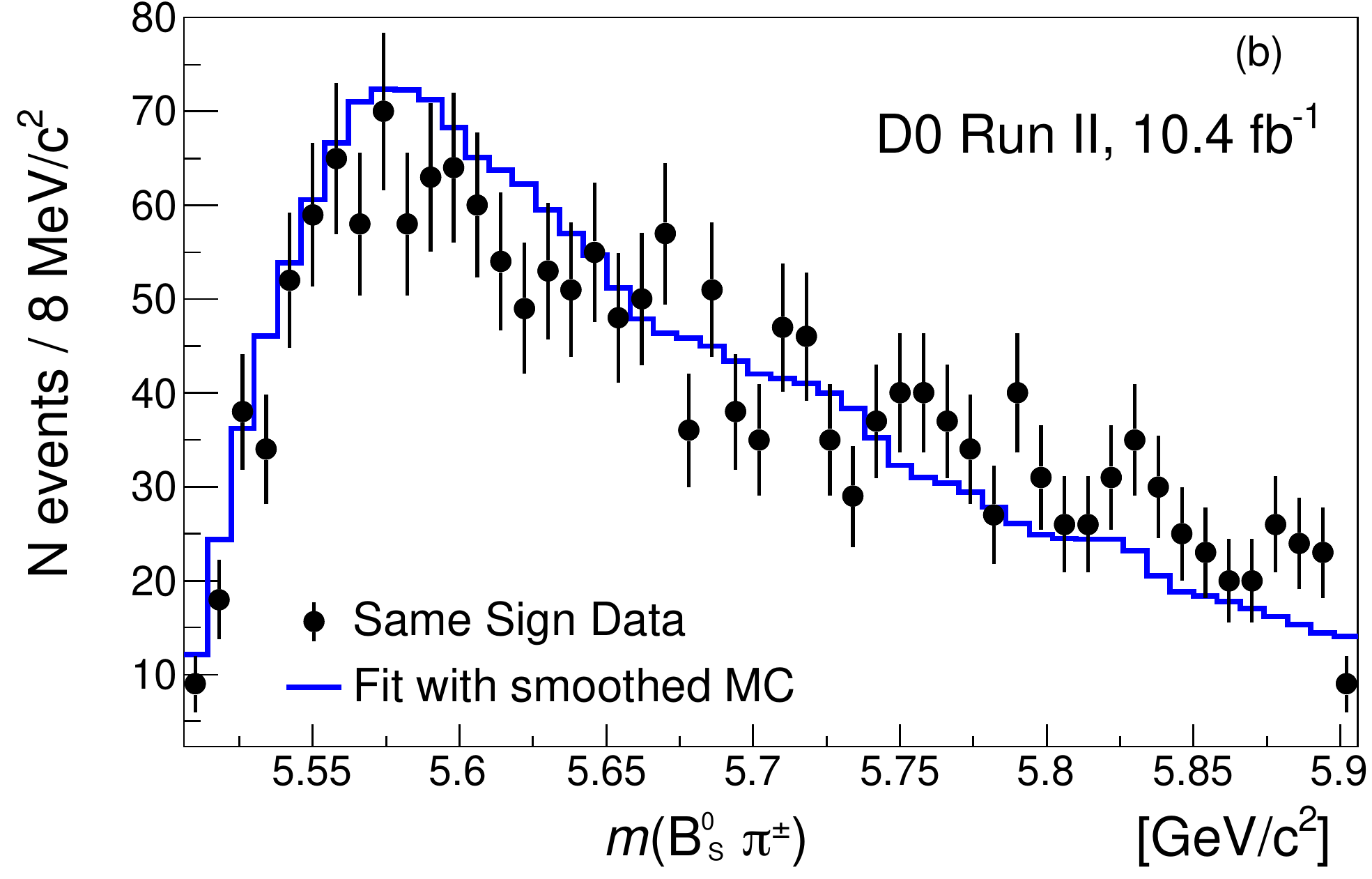}} 
     {\includegraphics[width=0.49\linewidth]{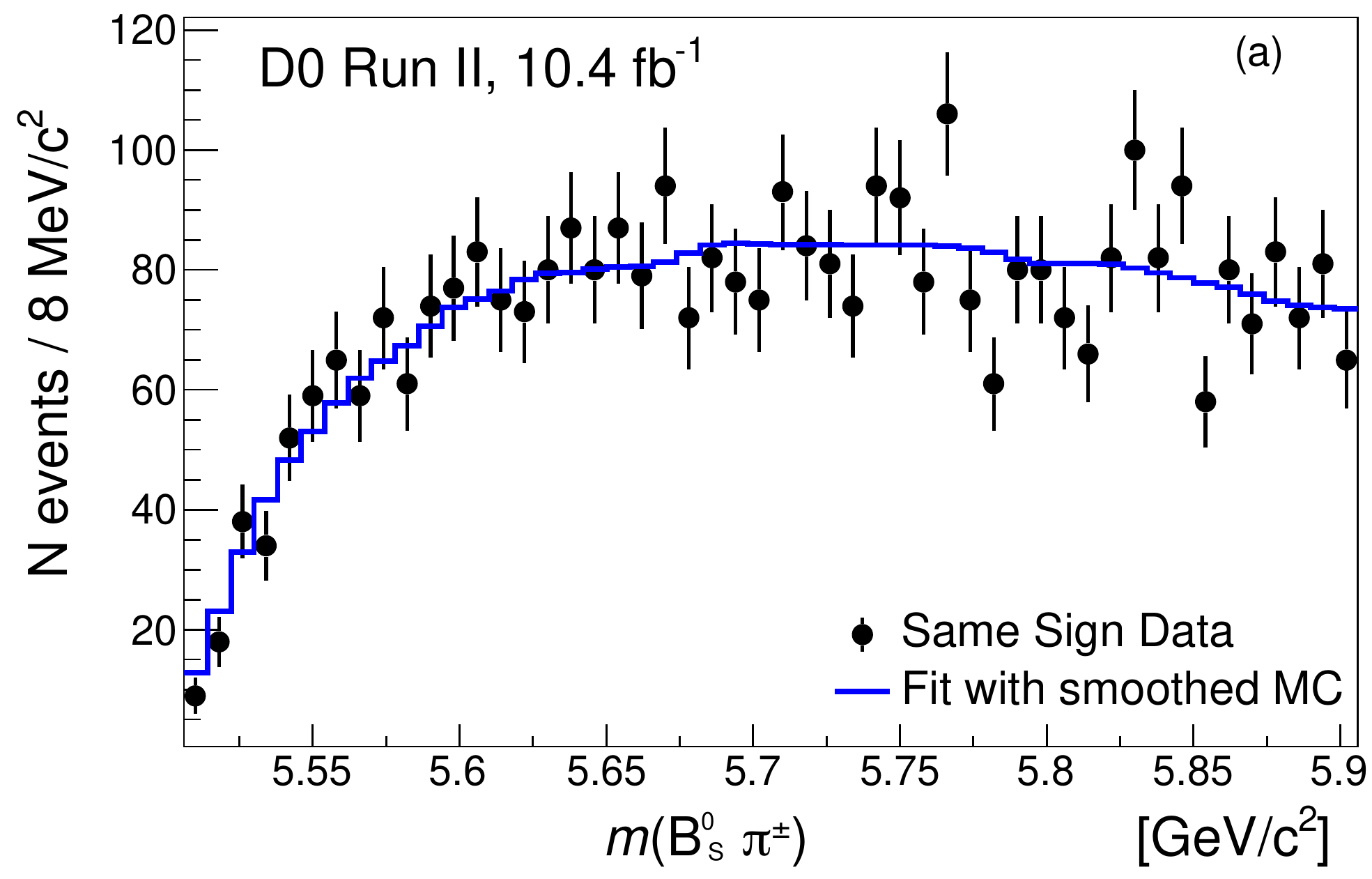}}
 
\caption{\label{fig:bkg_comp1} 
 The comparison of the $m(B_s^0 \pi^{\pm})$ background only distributions 
 a)  with the cone cut and b) without the cone cut, 
 obtained using the weighted MC (histogram) and from the same sign data samples (points with error bars). 
 The fluctuations in the number of MC events with the cone cut are due to the weighting
 procedure and the size of the sample. }
\end{figure*}

In Fig.~\ref{fig:bkg_comp1}(b) we compare the reweighted MC background simulation, smoothed using one iteration of
the 353QH algorithm~\cite{[{J. Friedman, }] [ {where for $n$ data points
the smoothed $i^{\mathrm{th}}$ point $y_s(i)$ is given by $y_s(i) =
0.25y(i-1) + 0.5y(i) + 0.25y(i+1)$ for $i = 2, n-1$ and $y_s(1) = y(1)$
and $y_s(n) = y(n)$.} ]Proceedings:1974sfa},  with 
the SS data for the no cone cut case.  
These two backgrounds are in good agreement since the $\chi^2$
between them is 50 for 50 bins.
We therefore choose to use the MC background shape only, for the data without the 
cone cut.  In  Fig.~\ref{fig:bkg_comp1}(a) we make the same comparison for the data 
with the cone cut.  In this case, $\chi^2$ = 77 for the 50 bins, and we therefore 
need to model the background shape with a combination of the MC and SS backgrounds.

To construct the background sample for the data with the cone cut the
fraction of MC and SS backgrounds need to be determined. 
This is found by fitting 
the data with a combination of the MC and SS with the fraction of MC events 
as a free parameter in the sideband mass range $5.506 < m(\Bs\pi^\pm) < 5.55$  and 
$5.650 < m(\Bs\pi^\pm) < 5.906$\,GeV/$c^2$. The best agreement is
found when the MC fraction is $\left(62 \pm 2\right)\%$. 

We choose the  background parametrization for the invariant mass distribution, 
both with and without the cone cut, to be
\begin{align}
\label{eq:ap_bkg}
   F_\mathrm{bgr} (m) = {}&  \left( C_1  m_0 + C_2  m_0^2 + C_3  m_0^3 + C_4  m_0^4 \right) \nonumber \\
   {}&  \times \exp \left( C_5  m_0 + C_7  m_0^2 \right),
\end{align}
where $m=m(B_s^0 \pi^{\pm})$,  $m_0 = m - m_\mathrm{th}$ and 
$m_\mathrm{th} = 5.5063$\,GeV$/c^2$ is the mass threshold. 
Our baseline choice of Eq.~(\ref{eq:ap_bkg}) gives an equivalently good description 
of the background as that used in Ref.~\cite{D0:2016mwd} [Eq.~(\ref{eq:ad_bkg})].  
It has the advantages of having one fewer parameter and being zero at the mass threshold.

Three alternative parametrizations are used to model the background.
The first is that used in Ref.~\cite{D0:2016mwd},
\begin{align}
 \label{eq:ad_bkg}
F_\mathrm{bgr} (m) = {}& \left( C_1 + C_2  m_\Delta^2 + C_3  m_\Delta^3 + C_4  m_\Delta^4 \right) \nonumber \\ 
   {}& \times  \exp \left( C_5 + C_6  m_\Delta + C_7  m_\Delta^2 \right) ,
\end{align}
where $m_\Delta = m - \Delta$ and $\Delta = 5.500$\,GeV$/c^2$. The 
second is the ARGUS function~\cite{Albrecht:1990am} which is
specifically constructed to describe background near a threshold
\begin{equation}
  \label{eq:ar_bkg}
F_\mathrm{bgr} (m) = m  \left( \frac{m^2}{m_\mathrm{th}^2} - 1 \right)^{C_1}  \exp \left( C_2  m\right).
\end{equation}
The third alternative model used to fit the background is  the MC
histogram (or combined MC and SS data) smoothed using one iteration of
the 353QH algorithm~\cite{[{J. Friedman, }] [ {where for $n$ data points
the smoothed $i^{\mathrm{th}}$ point $y_s(i)$ is given by $y_s(i) =
0.25y(i-1) + 0.5y(i) + 0.25y(i+1)$ for $i = 2, n-1$ and $y_s(1) = y(1)$
and $y_s(n) = y(n)$.} ]Proceedings:1974sfa}.

The ARGUS function is not used as an alternate parametrization in
the semileptonic data with the cone cut, because the fit to background 
is strongly disfavored
(the $\chi^2$ of the
fit to the MC background is 145 compared with approximately 50 for the
alternate functions). 
The $\chi^{2}$ per number of degrees of freedom ($\mathrm{ndf}$) for
the four representations of the background are shown in Table~\ref{tab:chi2}.

\begin{figure*}[htb]
 {\includegraphics[width=0.49\linewidth]{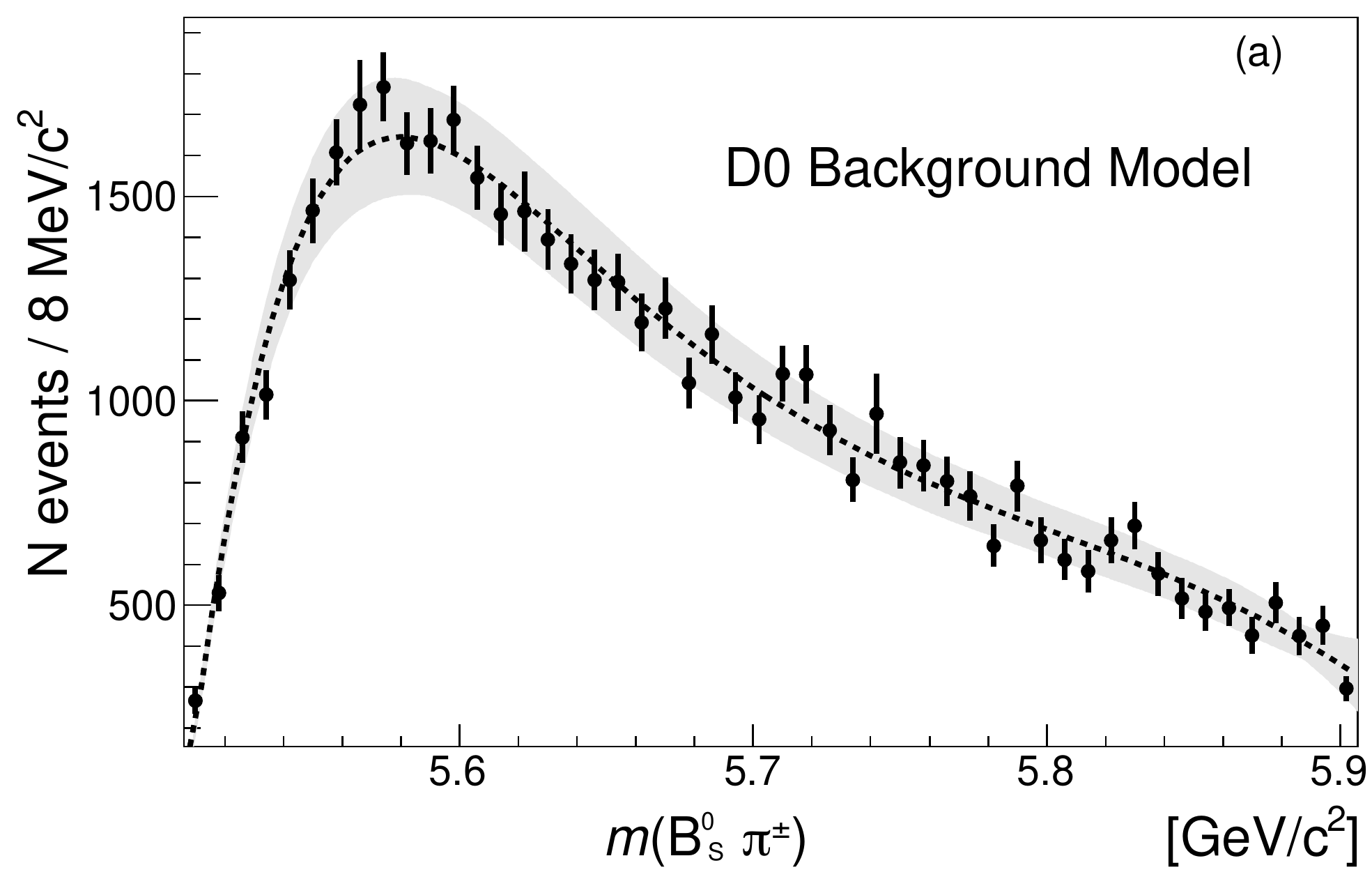}}  
 {\includegraphics[width=0.49\linewidth]{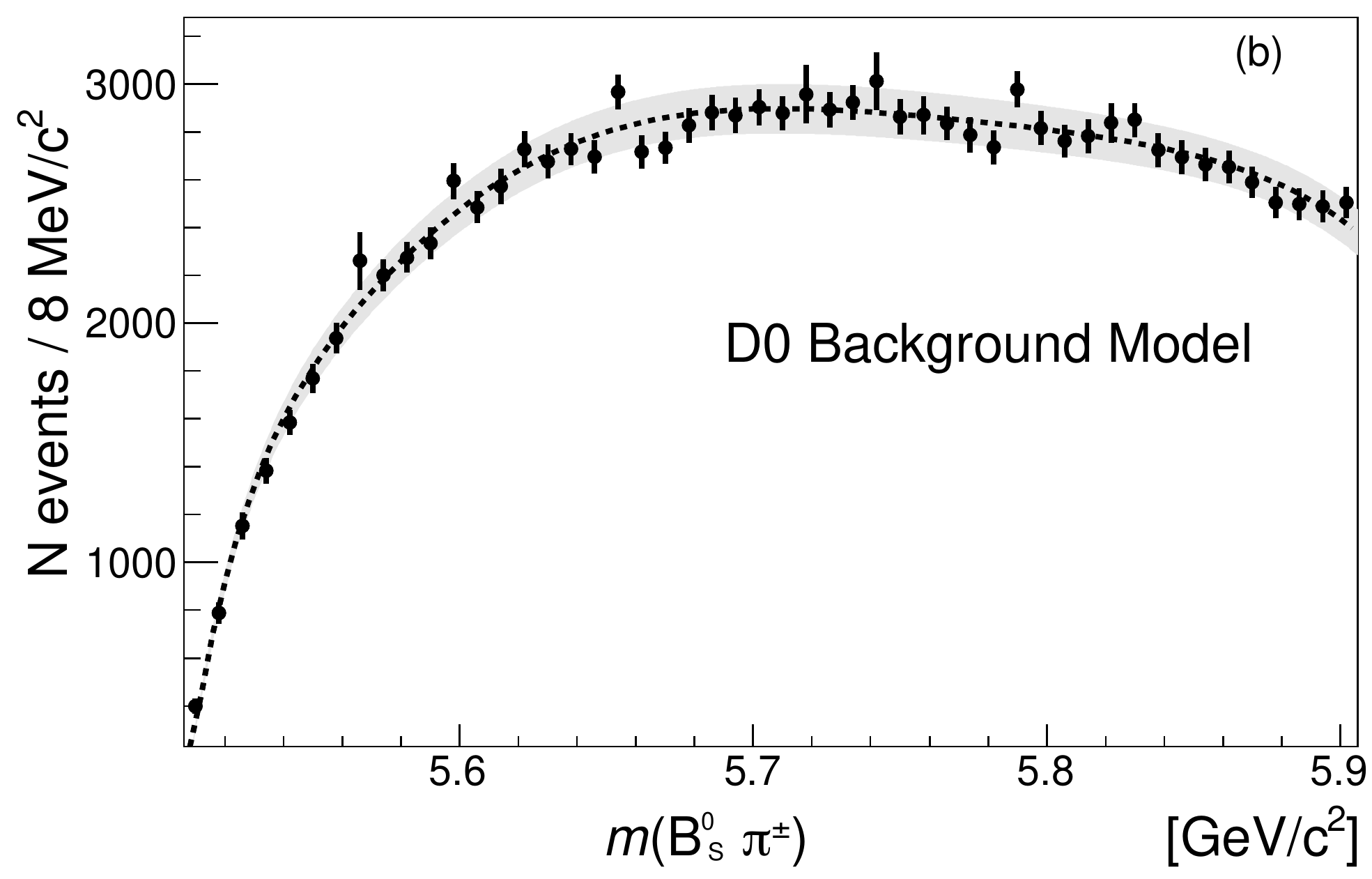}}
 \caption{\label{fig:bkg_comp} 
The background model produced according to the procedure described in
the text is shown along with background function~(\ref{eq:ap_bkg}) (dotted line) (a) with
and (b) without the cone cut. The gray band shows the systematic uncertainties on the background model (see Sec.~VI D).}
\end{figure*}

We choose the background description of Eq.~(\ref{eq:ap_bkg}) as the baseline.
 The alternative functions and the smoothed MC are used to
estimate the systematic uncertainty on the background shape. The
 $m(\Bs\pi^\pm)$  background model distribution along with  the
fit using Eq.~(\ref{eq:ap_bkg}) is presented in Fig.~\ref{fig:bkg_comp}.

\begin{table*}[bt]
\def\arraystretch{1.25}
\caption{\label{tab:chi2} Fit results for different parametrizations to the background model.}
\begin{ruledtabular}
\begin{tabular}{ccc}
Background function & \multicolumn{2}{c}{$\chi^{2}/$ndf }\\
& Cone cut & No cone cut \\
\hline 
Eq.~(1) 		& 51.0/(50-6) = 1.2 	& 48.1/(50-6) = 1.1 \\
Eq.~(2) 		& 42.9/(50-7) =	1.0	& 48.1/(50-7) = 1.1 \\
Eq.~(3) 		& 145/(50-2) =	3.0	& 38.3/(50-2) = 0.8 \\
Smoothed background 	& 33.8/(50-1) = 	0.7	& 30.9/(50-1) = 0.6\\
\end{tabular}
\end{ruledtabular}
\end{table*}

\section{Signal Mass Resolution}

We calculate the mass of the $\Bs\pi^\pm$ system using the quantity
\begin{equation}
 m(\Bs\pi^\pm) = m(\mu^\pm\Ds\pi^\pm) - m(\mu^\pm\Ds) + m(\Bs).
 \end{equation} 
Before carrying out the search for the $X^\pm(5568)$ in the semileptonic
channel we ensure that it is an unbiased and precise estimator of 
the mass of the $\Bs\pi^\pm$ system. 
This is studied by simulating the two
body decay $X(5568)^\pm \to \Bs\pi^\pm$ where \Bspmdecay, starting with a 
range of input masses
$\widetilde{m}(\Bs\pi^\pm)$.
Following the decay chain \Bspmdecay and
forming the invariant masses $m(\mu^\pm\Ds\pi^\pm)$ and $m(\mu^\pm\Ds)$
are found. 
Then   $m(\Bs\pi^\pm)$  is calculated and compared  to the input mass
$\widetilde{m}(\Bs\pi^\pm)$.

\begin{figure}[htbp]
\includegraphics[width=\linewidth]{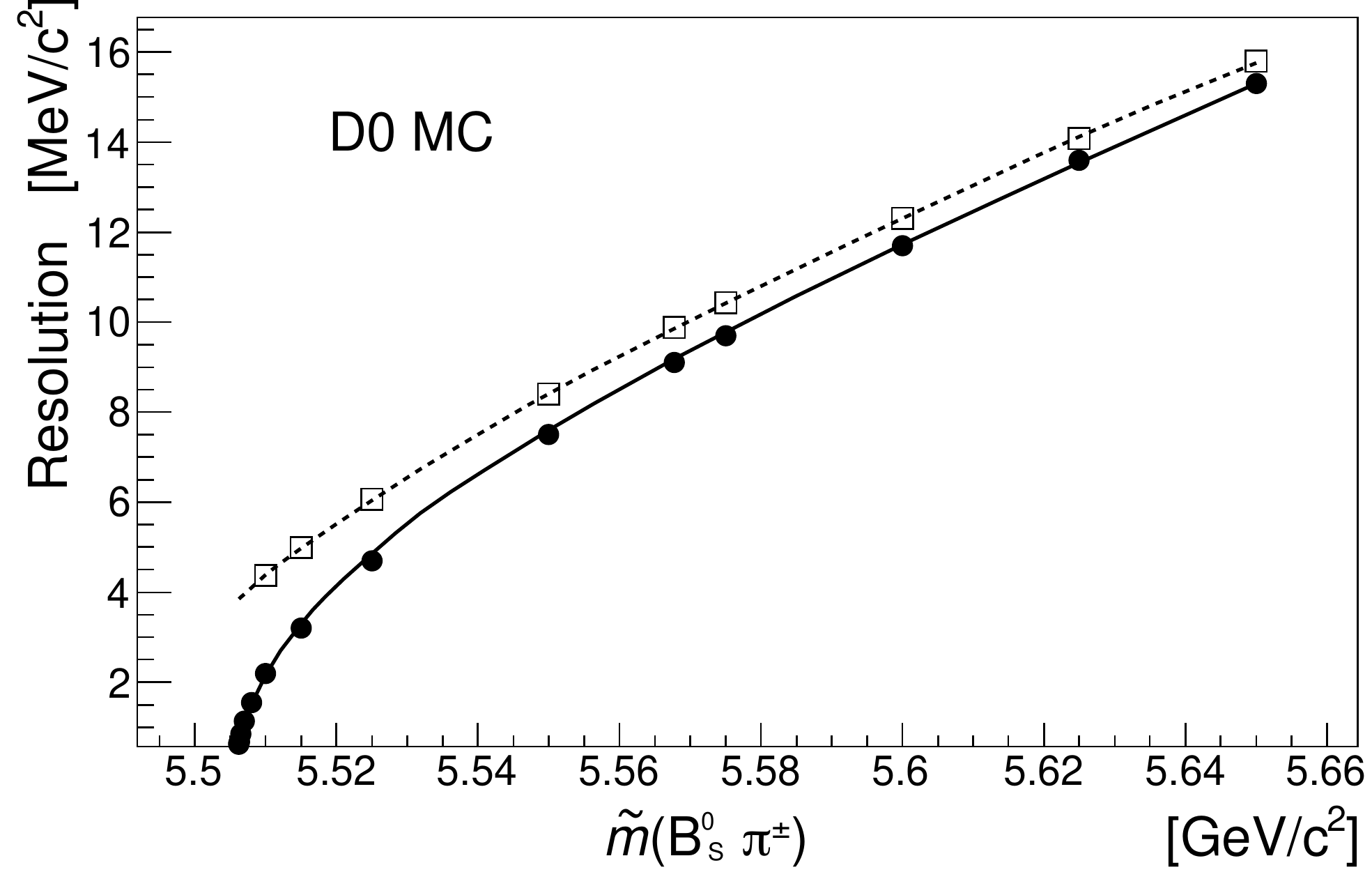}
\caption{\label{fig:SLres3} 
The resolution for a zero width resonance as a function of $\widetilde{m}(\Bs\pi^\pm)$. The solid
circles and the solid line show the  effect of the missing neutrino and
the open squares and dashed line show the convolution of the resolution
due to the missing neutrino convolved with the 3.85\,MeV/$c^2$ detector mass
resolution. 
 }
\end{figure}

To evaluate how well the mass approximation
works to compensate for the missing neutrino,
we model the decay with a toy MC that simulates the virtual $W$ in $\Bs \to
\Ds + W^\ast$ with an  isotropic distribution of $\mu$ and $\nu$ in the
$W$ boson rest frame.
The resulting resolution of a zero width resonance due to the presence of the neutrino  
is modeled by a Gaussian. 
The width varies according to  $\widetilde{m}(\Bs\pi^\pm)$ 
as illustrated by  the solid line in Fig.~\ref{fig:SLres3}.

The mass resolution for the D0 detector of a state decaying into five 
reconstructed charged particles with a similar kinematic range as in this 
study is measured using the MC simulation and is given by a Gaussian 
function of width 3.85\,MeV/$c^2$. The $m(\Bs\pi^\pm)$ resolution 
function is obtained by convoluting the Gaussian tracking resolution 
and the smearing resolution resulting from the missing neutrino.  
The resulting combined resolution, the dashed line in Fig.~\ref{fig:SLres3}, 
can be approximated by
\begin{equation}
\sigma_{\mathrm{SL}} = \left[ 3.85 + 60.93 (m_0^{0.85})  \right]  \,\mathrm{MeV}/c^2 
\label{eq:mass_resolution}
\end{equation}
where $m_0$ has the same definition as in Eq.~(\ref{eq:ap_bkg}).
These studies show that the difference between $m(\Bs\pi^\pm)$ and 
$\widetilde{m}(\Bs\pi^\pm)$ is less than  1\,MeV/$c^2$ in the 
search region. This is  confirmed  with the signal MC sample.

\section{Signal Fit Function}

The $X^\pm(5568)$ resonance is modeled by a relativistic Breit-Wigner function
convolved with a Gaussian detector resolution function given in 
Eq.~(\ref{eq:mass_resolution}), $F_{\rm sig}(m, m_X, \Gamma{_X})$, 
where $m_X$ and $\Gamma_X$ are the mass and the width of the resonance.

The fit function has the form
\begin{equation}
  \label{eq:spb}
   F = f_{\rm sig}  F_{\rm sig}(m, m_X, \Gamma{_X})  + f_{\rm bgr}  F_{\rm bgr} (m) ,
\end{equation}
where $f_{\rm sig}$ and $f_{\rm bgr}$ are normalization factors.  The
shape parameters in the background term $F_{\rm bgr}$ are fixed to the
values obtained from fitting the MC background distribution (see Fig.~\ref{fig:bkg_comp}).

We use the Breit-Wigner parametrization appropriate for an S-wave
two-body decay near threshold:
\begin{equation}
  \label{eq:signal_bw}
BW(m) \propto \frac{ m_X^2 \Gamma (m)}{(m_X^2 - m^2)^2 + m_X^2 \Gamma{^2} (m)}. 
\end{equation}
The mass-dependent width $\Gamma (m)=\Gamma_X \cdot (q_1 / q_0)$,  
where $q_1$ and $q_0$ are
the magnitudes of  momenta of the $B_s^0$ meson in the rest frame of the
$B_s^0 \pi^{\pm}$ system at the invariant mass equal to $m$ and $m_X$,
respectively.

\section{$\bm{X^\pm(5568)}$ semileptonic fit results}

\begin{figure*}[htbp]
{\includegraphics[width=0.49\linewidth]{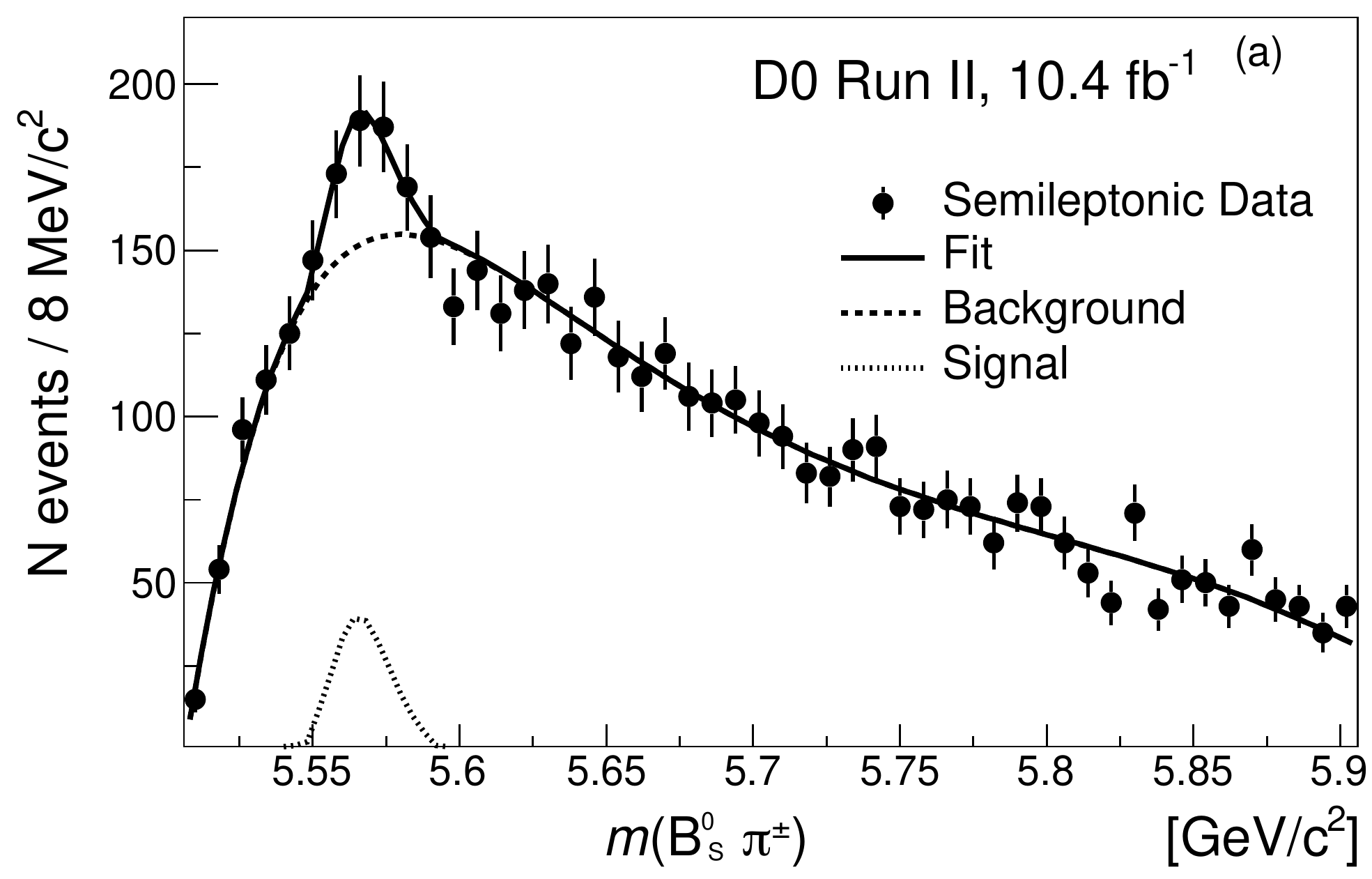}}
{\includegraphics[width=0.49\linewidth]{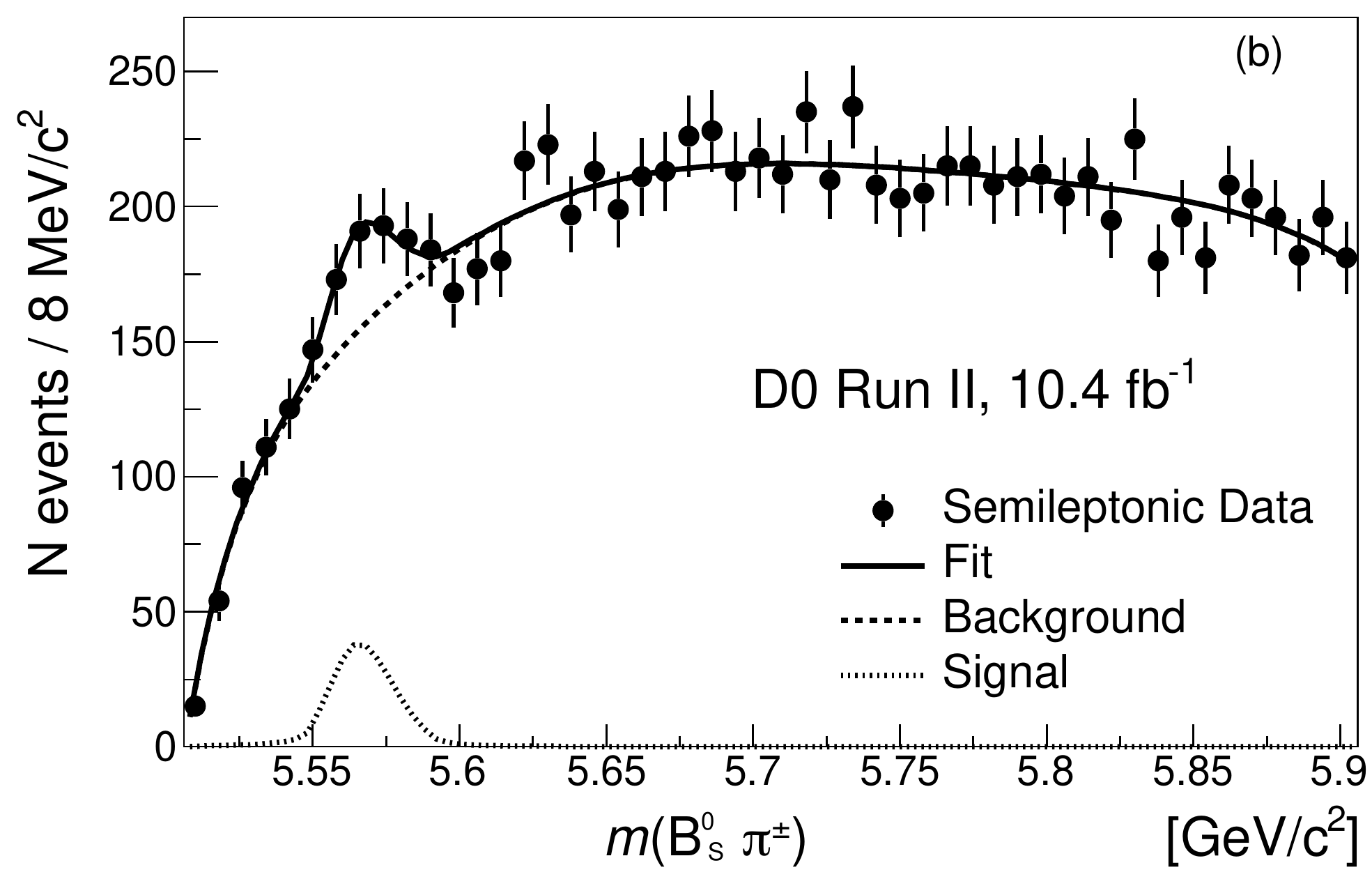}}
  \caption{\label{fig:fit_comp}  The $m(\Bs\pi^\pm)$ distribution (a) with
and (b) without the cone cut.  The
 fitting function is superimposed (see text for details).}
\end{figure*}

In the fit to the semileptonic data with the cone cut shown in
Fig.~\ref{fig:fit_comp}(a), the normalization parameters $f_{\rm sig}$ and
$f_{\rm bgr}$ and the Breit-Wigner parameters $m_X$ and $\Gamma{_X}$ are
allowed to vary. The fit yields the mass and width of $m_X =
5566.4^{+3.4}_{-2.8}$\,MeV/$c^2$, $\Gamma{_X} =
2.0^{+9.5}_{-2.0}$\,MeV/$c^2$, the number of signal events, $N =
121^{+51}_{-34}$, and a $\chi^2 = 34.9$ for 46 degrees of freedom. The
local statistical significance of the signal  is defined as \sig, where
${\cal{L}}_\mathrm{max}$ and ${\cal{L}}_0$ are likelihood values at the
best-fit signal yield and the signal yield fixed to zero obtained from a
binned maximum-likelihood fit. The $p$-value of the background only fit is $2.1 \times
10^{-5}$ and the local statistical significance is 4.3\,$\sigma$.

In the fit to the semileptonic data without the cone cut shown in
Fig.~\ref{fig:fit_comp}(b), the mass and width of $m_X =
5566.7^{+3.6}_{-3.4}$\,MeV/$c^2$, $\Gamma{_X} =
6.0^{+9.5}_{-6.0}$\,MeV/$c^2$, the number of signal events, $N =
139^{+51}_{-63}$, and a $\chi^2 = 30.4$ for 46 degrees of freedom.  The
$p$-value of the background only fit is $7.7 \times 10^{-6}$ and the local statistical
significance is 4.5\,$\sigma$. 
The fit results, both for the cone cut 
and no cone cut cases,  are given in Table~\ref{tab:tablesl} and for  various background parametrizations in Table~\ref{tab:table4}.  The $X^\pm(5568)$ parameters 
for the cone cut and no cone cut cases are consistent.

\begin{table*}[hbt]
\def\arraystretch{1.25}
\caption{\label{tab:tablesl} Results for the  fit to the semileptonic data sets (see Fig.~\ref{fig:fit_comp}).}
\begin{ruledtabular}
\begin{tabular}{ccc}
 & Cone cut & No cone cut \\
\hline
Fitted mass, MeV/$c^2$ 							& $5566.4^{+3.4}_{-2.8}\,\mbox{(stat)}^{+1.5}_{-0.6}\,\mbox{(syst)}$			& $5566.7^{+3.6}_{-3.4}\,\mbox{(stat)}^{+1.0}_{-1.0}\,\mbox{(syst)}$ \\
Fitted width, MeV/$c^2$ 						& $2.0^{+9.5}_{-2.0}\,\mbox{(stat)}^{+2.8}_{-2.0}\,\mbox{(syst)}$  				& $6.0^{+9.5}_{-6.0}\,\mbox{(stat)}^{+1.9}_{-4.6}\,\mbox{(syst)}$  	\\
Fitted number of signal events 					& $121^{+51}_{-34}\,\mbox{(stat)}^{+9}_{-28}\,\mbox{(syst)}$ 				& $139^{+51}_{-63}\,\mbox{(stat)}^{+11}_{-32}\,\mbox{(syst)}$ 	\\
$\chi^{2}/$ndf 									& $34.9/(50-4)$  																& $30.4/(50-4)$  			\\
$p$-value										& $2.1 \times 10^{-5}$     														& $7.7 \times 10^{-6}$      \\
Local significance 								& $4.3\, \sigma$  																& $4.5\, \sigma$  			\\
Significance including systematic uncertainties							& $3.2\, \sigma$  																& $3.4\, \sigma$  			\\
\end{tabular}
\end{ruledtabular}
\end{table*}

\begin{table*}[hbt]
\def\arraystretch{1.25}
\caption{\label{tab:table4} Semileptonic data fits for the different background parametrizations.}
\begin{ruledtabular}
\begin{tabular}{ccccc}
 & Equation (\ref{eq:ap_bkg})  & Equation (\ref{eq:ad_bkg})  & Equation (\ref{eq:ar_bkg}) & Smoothed MC simulation\\
\hline 
\multicolumn{5}{c}{Cone Cut}\\
\hline
Fitted mass, MeV/$c^2$ 				& $5566.4^{+3.4}_{-2.8} $ 		& $5566.1^{+3.7}_{-3.2}$ 	& \ldots\ 							& $5567.1^{+4.4}_{-3.3}$ \\
Fitted width, MeV/$c^2$ 			& $2.0^{+9.5}_{-2.0}$  			& $1.0^{+12.8}_{-1.0}$ 		& \ldots\ 							& $1.2^{+12.9}_{-1.2}$ \\
Fitted number of signal events 		& $121^{+51}_{-34}$ 			& $98^{+52}_{-29}$ 			& \ldots\ 							& $95^{+51}_{-30}$  \\
$\chi^{2}/$ndf 						& $34.9/(50-4)$  				& $43.2/(50-4)$ 			& \ldots\ 							& $50.5/(50-4) $  \\
Local significance 					& $4.3\, \sigma$  				& $3.6\, \sigma$ 			& \ldots\ 							& $3.5\, \sigma$ \\
\hline 
\multicolumn{5}{c}{No Cone Cut}\\
\hline
Fitted mass, MeV/$c^2$ 				& $5566.7^{+3.6}_{-3.4} $ 		& $5566.2^{+4.2}_{-4.1}$ 	& $5566.0^{+3.6}_{-3.4}$ 		& $5566.1^{+4.5}_{-4.5}$ \\
Fitted width, MeV/$c^2$ 			& $6.0^{+9.5}_{-6.0}$  			& $6.0^{+12.0}_{-6.0}$ 		& $6.5^{+8.9}_{-6.5}$ 			& $10^{+13}_{-10}$ \\
Fitted number of signal events 		& $139^{+51}_{-63}$ 			& $116^{+52}_{-48}$ 		& $146^{+51}_{-54}$ 			& $130^{+56}_{-48}$  \\
$\chi^{2}/$ndf 						& $30.4/(50-4)$  				& $50.3/(50-4)$ 			& $43.8/(50-4)$ 				& $44.8/(50-4) $  \\
Local significance 					& $4.5\, \sigma$  				& $3.7\, \sigma$ 			& $4.7\, \sigma$ 				& $3.8\, \sigma$ \\
\end{tabular}
\end{ruledtabular}
\end{table*}

\subsection{Systematic uncertainties}
\label{sec:systematic}

Systematic uncertainties (Table~\ref{tab:table1}) are obtained for the measured values of the
mass, width and event yield of the $X^\pm(5568)$ signal. The dominant
uncertainty is due to (i) the description of the background shape. We
evaluate this systematic uncertainty by using the alternative
paramaterizations of the background, Eqs. (2), (3) and the smoothed MC
histogram and finding the maximal deviations
from the nominal fit.

The effect of (ii) the MC weighting is estimated by creating 1000
background samples where the weights have been randomly varied based on
the uncertainties in the weighting procedure.

Other sources of systematic uncertainty are  evaluated by (iii) varying the
energy scale in the MC  sample relative to the data by
$\pm$1\,MeV/$c^2$,  (iv) varying the mass resolution of the $X^\pm(5568)$ 
either  by
$\pm$1\,MeV/$c^2$ around the mean value, or by  using a constant resolution of
11.1\,MeV/$c^2$ obtained from the MC simulation of the $X^\pm(5568)$ signal, (v) using a P-wave relativistic Breit-Wigner function, and
(vi) estimating the  shift  of the fitted mass peak due to  the missing
neutrino.

Systematic uncertainties  are summarized in Table~\ref{tab:table1}. The
uncertainties are added in quadrature separately for positive and
negative values to obtain the total systematic uncertainties for each
measured parameter. The results including systematic uncertainties are given in Table~\ref{tab:tablesl}.

\begin{table*}[htbp]
\def\arraystretch{1.25}
\caption{\label{tab:table1} Systematic uncertainties for the 
$X^\pm(5568)$ state mass,  width and the event yield obtained from the semileptonic data.  }
\begin{ruledtabular}
\begin{tabular}{lccc}
Source & Mass, MeV/$c^2$ & Width, MeV/$c^2$  & event yield, events  \\
\hline 
\multicolumn{4}{c}{ Cone Cut }\\
\hline
(i) Background shape description 			& +0.7 \ ;\ $-0.3$  & +0.0 \ ;\ $-1.0$ & +0.0 \ ;\ $-26.6$ \\
(ii) Background reweighting 				& +0.1 \ ;\ $-0.1$  & +0.4 \ ;\ $-0.4$ & +3.9 \ ;\ $-4.2$ \\
(iii) $B_s^0$ mass scale, MC simulation and data 		& +0.1 \ ;\ $-0.3$  & +0.8 \ ;\ $-1.0$ & +5.1 \ ;\ $-7.8$ \\
(iv) Detector resolution 					& +0.9 \ ;\ $-0.0$ 	& +2.7 \ ;\ $-1.0$ & +6.5 \ ;\ $-0.0$ \\
(v) {\it P}-wave Breit-Wigner  				& +0.0 \ ;\ $-0.4$ 	& +0.0 \ ;\ $-1.0$ & +0.0 \ ;\ $-3.7$ \\
(vi) Missing neutrino effect  				& +1.0 \ ;\ $-0.0$ 	& - & - \\
\hline
Total 									& +1.5 \ ;\ $-0.6$ 	& +2.8 \ ;\ $-2.0$ & +9.1 \ ;\ $-28.3$ \\
\hline
\multicolumn{4}{c}{No Cone Cut}\\
\hline
(i) Background shape description 			& +0.0 \ ;\ $-0.7$  & +0.7 \ ;\ $-2.5$ & +4.8 \ ;\ $-28.0$ \\
(ii) Background reweighting 				& +0.1 \ ;\ $-0.1$  & +0.7 \ ;\ $-0.7$ & +5.0 \ ;\ $-5.0$ \\
(iii) $B_s^0$ mass scale, MC simulation and data 		& +0.3 \ ;\ $-0.5$ 	& +1.0 \ ;\ $-1.4$ & +7.5 \ ;\ $-9.6$ \\
(iv) Detector resolution 					& +0.0 \ ;\ $-0.5$ 	& +1.3 \ ;\ $-2.6$ & +3.7 \ ;\ $-6.4$ \\
(v) {\it P}-wave Breit-Wigner  				& +0.0 \ ;\ $-0.2$ 	& +0.0 \ ;\ $-2.4$ & +0.0 \ ;\ $-7.0$ \\
(vi) Missing neutrino effect  				& +1.0 \ ;\ $-0.0$ 	& - & - \\
\hline
Total 									& +1.0 \ ;\ $-1.0$ & +1.9 \ ;\ $-4.6$ & +10.9 \ ;\ $-31.5$ \\
\end{tabular}
\end{ruledtabular}
\end{table*}

\subsection{Significance}

Since we are seeking to
confirm the result presented in Ref.~\cite{D0:2016mwd} we do not apply a
correction for a look elsewhere effect.
The systematic uncertainties are treated as nuisance parameters to
construct a prior predictive model~\cite{Olive:2016xmw,Giunti:1998xv} of
our test statistic. When the systematic uncertainties are included, the
significance of the observed semileptonic signal with the cone cut is {
3.2\,$\sigma$ ($p$-value = $1.4 \times 10^{-3}$)}. The
significance of the  semileptonic signal without the cone is 
3.4\,$\sigma$ ($p$-value = $6.4 \times 10^{-4}$). 

\subsection{Closure tests}
\label{sec:closure1}

\begin{figure}[htb]
  {\includegraphics[width=\linewidth]{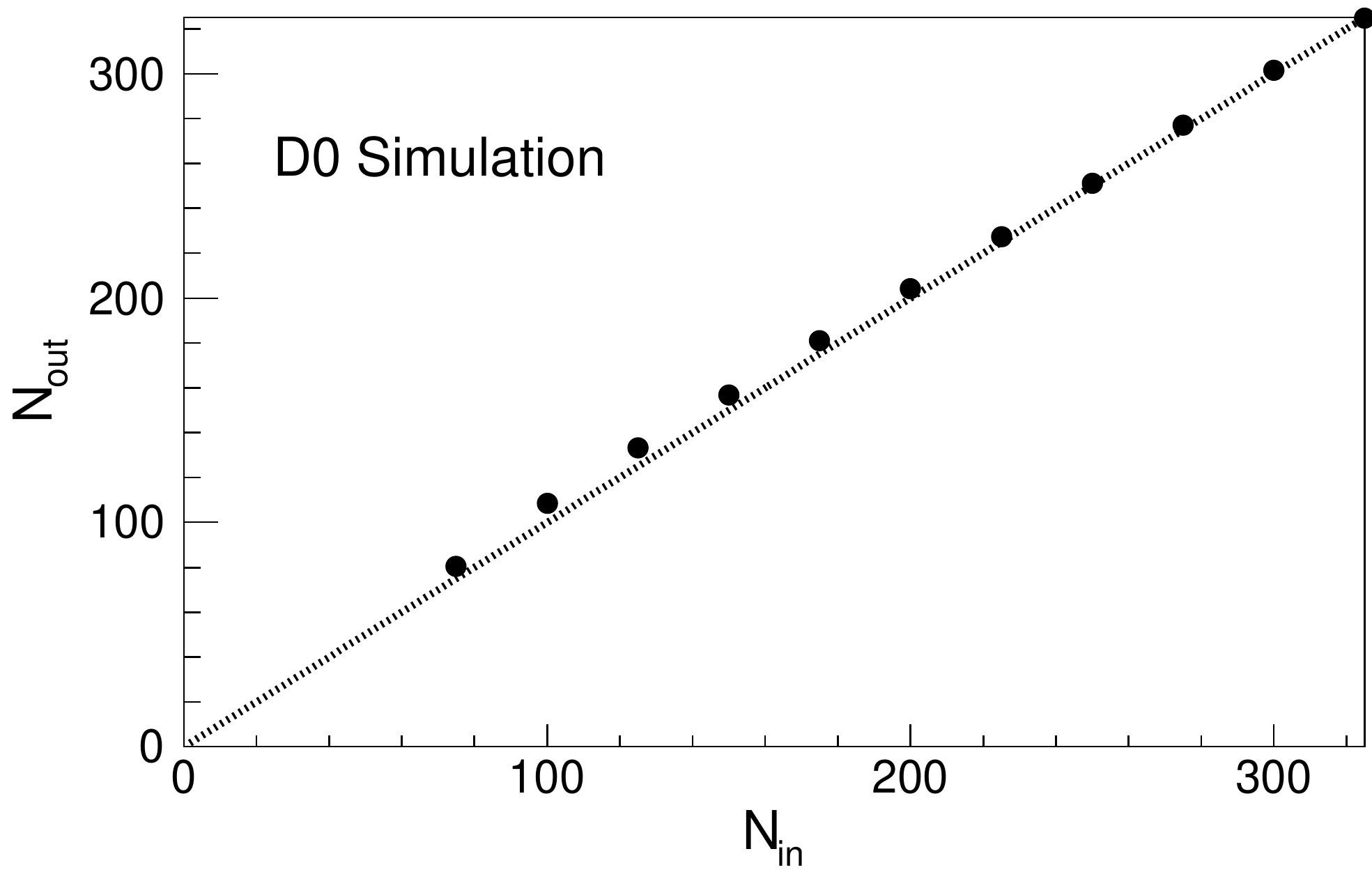}}
  \caption{\label{fig:si_plot_sl} Results of the toy MC tests of the  fitting
procedure (black circles) used in the analysis of the semileptonic data with   the cone cut.  
The number of fitted signal events
 are plotted vs    fitted number of injected
signal events. The dotted line shows
$N_\mathrm{in}=N_\mathrm{out}$.}
\end{figure}

We have tested the accuracy of the fitting procedure using toy MC event 
samples constructed with input mass and width of 5568.3 and 21.9\,MeV$/c^2$ 
respectively, with the number of input signal events varied in steps of 25 
between 75 and 350. At each number of input signal events, 10,000 
pseudoexperiments were generated.
The signals are modeled with a relativistic Breit-Wigner
function convolved with a Gaussian function  representing the appropriate detector
resolution. 
The background distribution is based on Eq.~(\ref{eq:ap_bkg}).
For each trial the fitting procedure is
performed to obtain the  mass and width and the number of semileptonic signal events.
The results of each set of trials is fitted with a Gaussian to determine
the mean and the uncertainty in the number of signal events, the mass
and the width (see Table~\ref{tab:tabletoyMCSL}). The number of fitted
signal events vs. the number of injected signal events for the semileptonic 
 samples are plotted in Fig.~\ref{fig:si_plot_sl}.

For the ensembles with a number of input events similar to that observed in data, 
there is a slight overestimate of the yield and fitted mass, and the width is 
underestimated.   This width reduction is in agreement with the results of the fits 
to data (Sec.~VII), and indicate that the semileptonic data are not sensitive to the width.
These effects are accounted for in the calculation of the significance.

\begin{table*}[hbtp]
\def\arraystretch{1.25}
\caption{\label{tab:tabletoyMCSL} 
 Mean values and uncertainties for fitted number of events, 
 mass and width 
 from
 Gaussian fits to corresponding distributions from 10,000 trials with the cone cut. 
 Also given is the expected statistical uncertainties on the fitted number of events, $\Delta$($N_{\mathrm{fit}}(\mathrm{sl})$),
 and the expected uncertainties on the measurement of the width, $\Delta(\Gamma_X)$\,MeV/$c^2$. 
  A range
 of signals  with 75, 100, 125, 150, 175 and 200 signal events, mass $m_x
 = 5568.3$\,MeV$/c^2$ and width $\Gamma_X = 21.9$\,MeV$/c^2$ have been
 simulated. Background parameterization Eq.~(\ref{eq:ap_bkg}) is used.}
\begin{ruledtabular}
\begin{tabular}{cccccccccc}
$N_{\mathrm{in}}(\mathrm{sl})$ & $N_{\mathrm{fit}}(\mathrm{sl})$ & $\Delta$($N_{\mathrm{fit}}(\mathrm{sl})$) &
$m_X$ MeV/$c^2$ & $\Gamma_X$ MeV/$c^2$ & $\Delta(\Gamma_X)$ MeV/$c^2$ \\
\hline
75	& $80.4  \pm	0.9$ &	61 & 	$5577.9 \pm	0.24$	& 13.1	& 15.3  \\
100	& $108.5 \pm	0.7$ &	58 & 	$5572.9 \pm	0.17$	& 15.8	& 15.6  \\
125	& $133.3 \pm	0.6$ &	59 & 	$5570.4 \pm	0.12$	& 17.7	& 15.3  \\
150	& $156.7 \pm	0.6$ &	58 & 	$5569.3 \pm	0.08$	& 19.3	& 14.6  \\
175	& $181.0 \pm	0.6$ &	59 & 	$5568.9 \pm	0.07$	& 20.2	& 13.8  \\
200	& $204.2 \pm	0.6$ &	61 & 	$5568.7 \pm	0.05$	& 20.8	& 12.9  \\
\end{tabular}
\end{ruledtabular}
\end{table*}

\subsection{Comparison with hadronic channel}

The measured values of the mass, width, the number of signal events, and
significance of the signal for the semileptonic channel and the hadronic
channel~\cite{D0:2016mwd} 
are given in Table~\ref{tab:table4a}. 
The mass and width of the $X^\pm(5568)$  for the 
semileptonic and hadronic channels are consistent taking into account  the uncertainties. 
The observed yields are consistent with coming from a common particle given the number of 
$B_s^0$ events in the sample  and the \Bs\ branching ratios.

\begin{table*}[bpht]
\def\arraystretch{1.25}
\caption{\label{tab:table4a} Fit results obtained in the semileptonic channel
and in the hadronic channel (Ref.~\cite{D0:2016mwd}). In the hadronic channel with no cone cut 
the mass and  width of the $X^\pm(5568)$ were set to the values found with the cone cut. 
LEE - look elsewhere effect.}
\begin{ruledtabular}
\begin{tabular}{lcccc}
 & \multicolumn{2}{c}{Semileptonic} & \multicolumn{2}{c}{Hadronic (from Ref.~\cite{D0:2016mwd})} \\
 & Cone cut  & No cone cut &  Cone cut & No cone cut  \\
\hline
Fitted mass, MeV/$c^2$ 				& $5566.4^{+3.4}_{-2.8}\ ^{+1.5}_{-0.6}$	& $5566.7^{+3.6}_{-3.4}\ ^{+1.0}_{-1.0}$	& $5567.8 \pm 2.9 ^{+0.9}_{-1.9}$ 		& 5567.8 \\
Fitted  width, MeV/$c^2$ 	        & $2.0^{+9.5}_{-2.0}\ ^{+2.8}_{-2.0}$ 		& $6.0^{+9.5}_{-6.0}\ ^{+1.9}_{-4.6}$ 		& $21.9 \pm 6.4 ^{+5.0}_{-2.5}$ 		& 21.9 \\
Fitted number of signal events 		& $121^{+51}_{-34}\ ^{+9}_{-28}$ 			& $139^{+51}_{-63}\ ^{+11}_{-32}$ 			& $133 \pm 31 \pm 15$ 					& $106 \pm 23 {\rm
\thinspace (stat)}$ \\
Local significance 					& $4.3\, \sigma$ 							& $4.5\, \sigma$ 							& $6.6\,\sigma$ 							& $4.8\,\sigma$ \\
Significance with systematics 		& $3.2\, \sigma$ 							& $3.4\, \sigma$ 							& $5.6\, \sigma$ 							& \ldots\ \\
Significance with LEE+systematics 	& \ldots\										& \ldots\										& $5.1\,\sigma$ 							& $3.9\,\sigma$ \\
\end{tabular}
\end{ruledtabular}
\end{table*}

\subsection{Cross-checks}

As a cross-check  the  $\Bs\pi^\pm$ mass-bin size is set to  5\,MeV/$c^2$ and to
10\,MeV/$c^2$ instead of 8\,MeV/$c^2$, and  the lower edge of
the fitted mass range is shifted by 2, 3, 5, and 7\,MeV/$c^2$. This leads
to maximal variations in the mass of $^{+0.1}_{-0.6}$\,MeV/$c^2$, in the
 width of $^{+1.7}_{-0.9}$\,MeV/$c^2$ and in the number of signal
events $^{+0}_{-9}$ which are  small compared to the statistical and
systematic uncertainties.

To test the stability of the results, alternative choices are made
regarding the fit parameters. In the first, the background fit
parameters are allowed to float. The resulting fit  is  consistent with
the nominal fit and  the  $p$-value of the background-only fit is  $1.7 \times 10^{-4}$ corresponding
to a statistical significance of 3.8\,$\sigma$
(Table~\ref{tab:fitCrossChecks}). The second cross-check fixes the mass
and  width of the $X^\pm(5568)$ to the values found in
Ref.~\cite{D0:2016mwd}. Again, the resulting fit  is  consistent with
the nominal fit with an increase in the number of signal events due to the increased
width of the peak.  The $p$-value of the background-only fit is $1.1 \times 10^{-4}$
corresponding to a statistical significance of 3.9\,$\sigma$
(Table~\ref{tab:fitCrossChecks}). These cross-checks are also repeated without the cone cut (Table~\ref{tab:fitCrossChecks}).

\begin{table*}[htbp]
\def\arraystretch{1.25}
\caption{\label{tab:fitCrossChecks}  Fit results for the semileptonic channel using parametrization 
(\ref{eq:ap_bkg}) with the nominal fit, with all parameters free and the mass and width fixed to those 
of the hadronic channel. Statistical uncertainties only.   }
\begin{ruledtabular}
\begin{tabular}{cccc}
 									& Nominal fit 				&  All parameters free  	& Mass and width fixed to hadronic  \\
\hline 
\multicolumn{4}{c}{ Cone Cut  }\\
\hline
Fitted mass, MeV/$c^2$ 				& $5566.4^{+3.4}_{-2.8}$ 	& $5567.2\pm 2.9$ 			& $5567.8$ 	 \\
Fitted width, MeV/$c^2$ 			& $2.0^{+9.5}_{-2.0}$  		& $8.3\pm11.0$ 				& $21.9$  		 \\
Fitted number of signal events 		& $121^{+51}_{-34}$ 		& $181\pm88$ 				& $164 \pm 44$ 	  \\
$\chi^{2}/$ndf 						& $34.9/(50-4)$  			& $30.9/(50-10)$ 			& $38.0/(50-2)$  			 \\
Local significance 					& $4.3 \sigma$  			& $3.8 \sigma$ 				& $3.9 \sigma$  			\\
\hline 
\multicolumn{4}{c}{ No cone cut  }\\
\hline
Fitted mass, MeV/$c^2$ 				& $5566.7^{+3.6}_{-3.4} $ 	& $5566.6\pm 3.5$ 			& $5567.8$ 	 \\
Fitted width, MeV/$c^2$ 			& $6.0^{+9.5}_{-6.0}$  		& $8.4\pm14.5$ 				& $21.9$  		 \\
Fitted number of signal events 		& $139^{+51}_{-63}$ 		& $144\pm101$ 				& $168 \pm 42$ 	  \\
$\chi^{2}/$ndf 						& $30.4/(50-4)$  			& $27.4/(50-10)$ 			& $32.8/(50-2)$  			 \\
Local significance 					& $4.5 \sigma$  			& $4.4 \sigma$ 				& $4.2 \sigma$  			\\
\end{tabular}
\end{ruledtabular}
\end{table*}

\section{Production ratio of $\bm{X^\pm(5568)}$ to $\bm{B_s^0}$}
\label{sec:ratio}

To calculate the production ratio of the $X^\pm(5568)$ to $B_s^0$, the number
of the $B_s^0$-mesons needs to be estimated. The fitting of the $K^+K^-\pi^\mp$  mass distribution 
is described in Sec.~\ref{sec:background}.
The number of \Ds\ mesons
extracted from the fit and adjusted for the mass range $1.91 <
m(K^+K^-\pi^\mp) < 2.03$\,MeV$/c^2$ is $N(\Ds) = 6648 \pm 127$ (see
Fig.~\ref{fig:ad3}).
The number of $\mu^\pm\Ds$ events in the signal sample that are the
result of a random combination of a promptly produced \Ds\ and a muon
in the event is estimated using events where the muon and the \Ds-meson
have the same sign. The same sign data sample is analyzed using the same
model as the opposite sign data with the means and widths of the
Gaussians fixed to the values obtained from the opposite sign data. The
number of events in the same-sign sample is $N(D_s^\pm) = 426 \pm 61$.
The mass distributions of the $K^+K^-\pi^\mp$ for opposite and same-sign
data are shown in Fig.~\ref{fig:ad3}.

The number of \Bs-meson decays in the semileptonic  data is estimated by subtracting 
the contribution of the promptly produced $\mu^\pm\Ds$ events from the overall   
$\mu^\pm\Ds$ sample.
A study of the MC background simulations shows that the purity of the resulting 
sample is $99.5^{+0.5}_{-1.0}\%$. We find  $6222 \pm 141$ \Bs\ events.

Combining these results and using the efficiency for the charged pion 
in the X(5568) decay (Sec. IV), we obtain
 a  production ratio for the
semileptonic data of  
\begin{align}
\rho &{}= \frac{N_\mathrm{sl}(X^\pm(5568)\rightarrow \Bs(\mathrm{sl}) \pi^\pm)}{N_{\Bs}(\mathrm{sl})} \nonumber \\ 
&{} = \left[7.3^{+2.8}_{-2.4} \thinspace  {\rm
(stat)} ^{+0.6}_{-1.7} \thinspace  {\rm (syst)}\right]\%,
\end{align}
 for  our
fiducial selection (which includes the requirements 
$p_T(\mu^\pm\Ds)>10\,$GeV$/c^2$ and  $4.5\,$GeV$/c^2\, < m(\mu^\pm \Ds) <
m(\Bs)$), where $N_\mathrm{sl}(X^\pm(5568)\rightarrow \Bs(\mathrm{sl}) \pi^\pm)$ 
is the number of $X^\pm(5568)$ decays to $\Bs \pi^\pm$ and 
$N_{\Bs}(\mathrm{sl})$ is the inclusive number of \Bs\ decays, both for semileptonic decays of the \Bs .
This result is similar to the ratio 
measured in the hadronic
channel $\left[8.6 \pm 1.9 \thinspace  {\rm (stat)} \pm 1.4 \thinspace 
{\rm (syst)}\right]\%$ for 
$p_T(J/\psi\phi\pi^\pm)>10\,$GeV$/c^2$~\cite{D0:2016mwd}.

\section{Combined  Signal Extraction}

We now proceed to fit the hadronic and semileptonic data sets simultaneously.
The hadronic data set is the same as used in Ref.~\cite{D0:2016mwd}
except that the data are fitted in the mass range $5.506 <
m(\Bs\pi^\pm) < 5.906$\,GeV/$c^2$ instead of $5.500 < m(\Bs\pi^\pm) <
5.900$\,GeV/$c^2$. 
The data selection and background modeling for the hadronic data set are
described in detail in Ref.~\cite{D0:2016mwd}.

\begin{figure*}[hbt]
  {\includegraphics[width=0.49\linewidth]{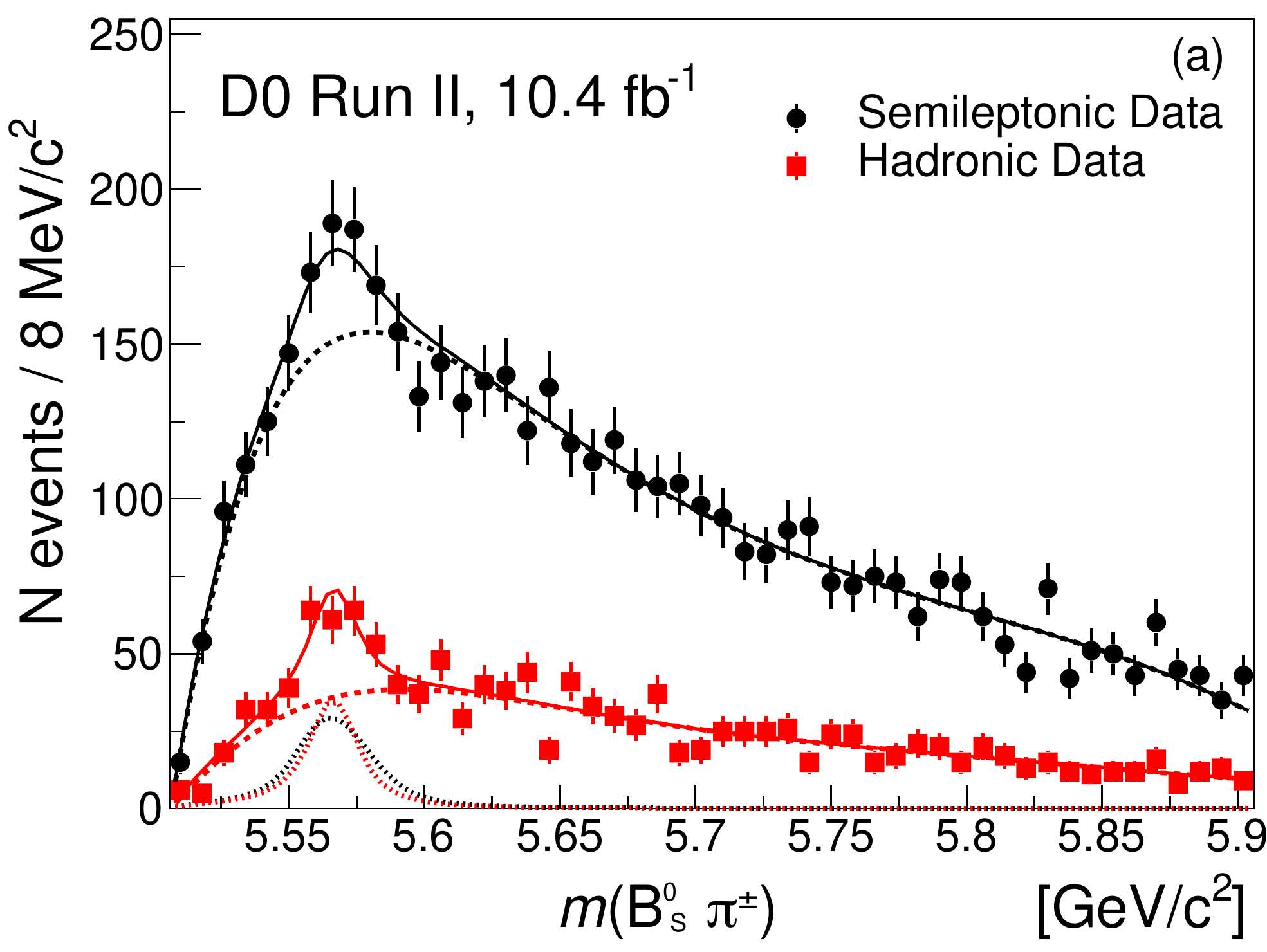}}
  {\includegraphics[width=0.49\linewidth]{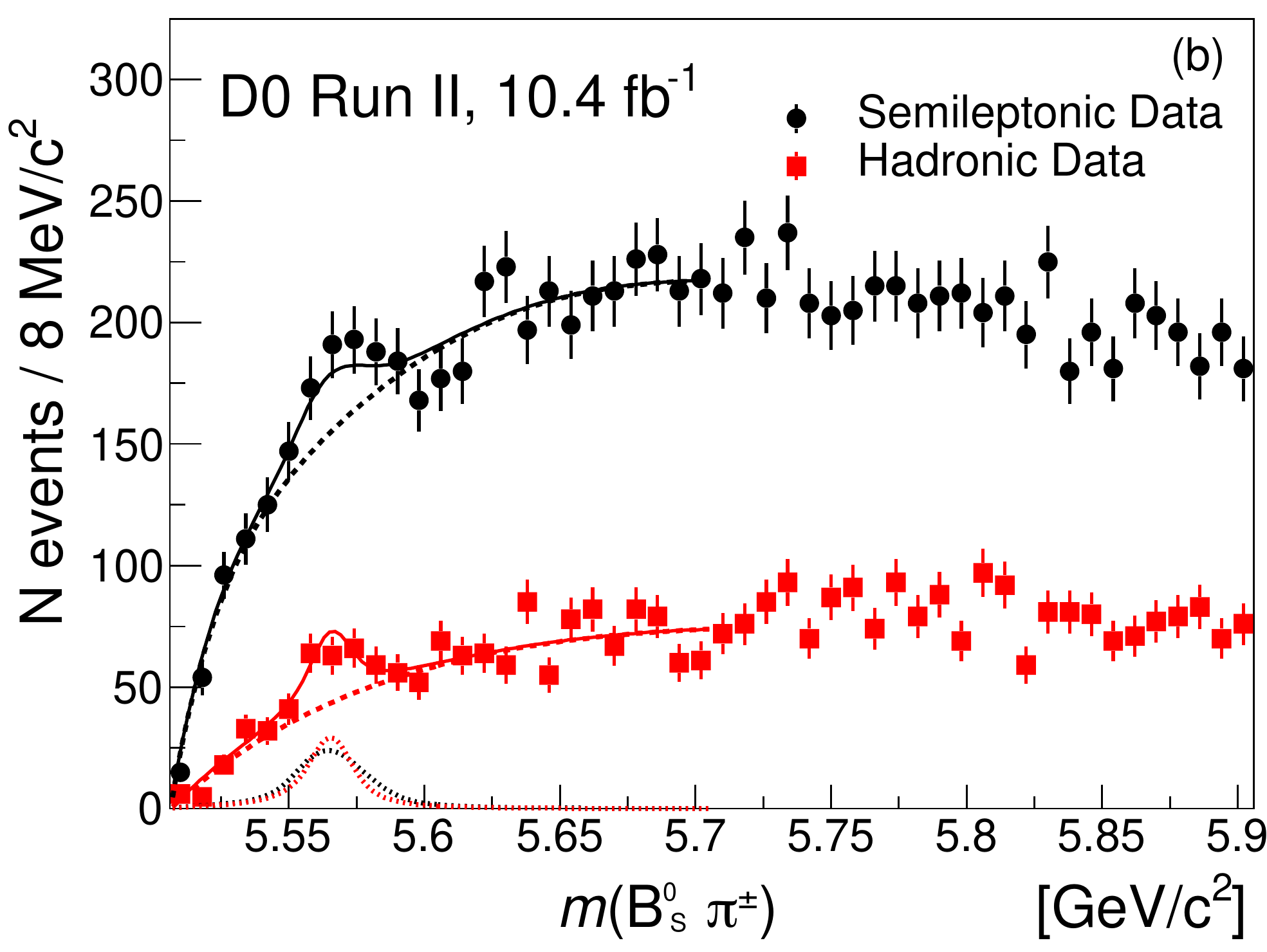}}
\caption{\label{fig:fit_combined} The $m(\Bs\pi^\pm)$ distribution for the hadronic (red squares) and semileptonic (black circles) 
data   with the combined fitting function superimposed (a) with and (b) without the cone cut.
(see text for details, the resulting fit parameters are given in Table~\ref{tab:tablecombined}).
The background parametrization function is taken from Eq.~(\ref{eq:ap_bkg}). }
\end{figure*}

The fit function has the form
\begin{align}
  \label{eq:combined}
   F_{\rm h} = &{} f_{\rm h,sig}  F_{\rm h,sig}(m, m_X, \Gamma{_X})  + f_{\rm h,bgr}  F_{\rm h,bgr} (m) ,\\
   F_{\rm sl} = &{} f_{\rm sl,sig}  F_{\rm sl,sig}(m, m_X, \Gamma{_X})  + f_{\rm sl,bgr}  F_{\rm sl,bgr} (m) 
\end{align}
where $f_{\rm h(sl),sig}$ and $f_{\rm h(sl),bgr}$ are
normalization factors.  The shape parameters in the background terms
$F_{\rm h(sl),bgr}$ are fixed to the values obtained from fitting the
respective background models  for the hadronic (h) and semileptonic (sl)
samples  to Eq.~(\ref{eq:ap_bkg}).  The signal
shape $F_{\rm h(sl),sig}$ is modeled by
relativistic Breit-Wigner function convolved with a Gaussian detector
resolution function that depends on the data sample. For the
semileptonic sample the detector resolution is given by
Eq.~(\ref{eq:mass_resolution}) and for the hadronic channel it is
3.85\,MeV/$c^2$. For the data without the cone cut the combined data are
fitted in the range $5.506 < m(\Bs\pi^\pm) < 5.706$\,GeV/$c^2$ as the
hadronic  background is not well modeled for $m(\Bs\pi^\pm) >
5.706$\,GeV/$c^2$~\cite{D0:2016mwd}. The same Breit-Wigner parameters
$m_X$ and $\Gamma_{X}$ are used for the hadronic and semileptonic
samples. In the fits shown in Fig.~\ref{fig:fit_combined}, the
normalization parameters $f_{\rm h(sl),sig}$ and $f_{\rm h(sl),bgr}$ and
the Breit-Wigner parameters $m_X$ and $\Gamma{_X}$ are allowed to vary.
Since the fraction of \Bs\ events produced by the decay of the $X^\pm(5568)$
should be essentially the same in the hadronic and semileptonic channels 
the $X^\pm(5568)$ event yields ($N_{\mathrm h}$ and $N_{\mathrm{sl}}$) are constrained using the parameter 
\begin{equation}
A_{\mathrm{sl,h}} = \frac{N_{\mathrm{sl}} - N_{\mathrm h}}{N_{\mathrm{sl}} + N_{\mathrm h}} 
\label{eq:prodfrac}
\end{equation}
which is required to be consistent with the \Bs-meson production rate in the hadronic and semileptonic channels 
\begin{equation}
A_{\mathrm{sl,h}}(\Bs) = 
\frac{N_{\Bs}(\mathrm{sl}) - N_{\Bs}(\mathrm h)}{N_{\Bs}(\mathrm{sl}) + N_{\Bs}(\mathrm h)} = 0.054 \pm 0.020,
\label{eq:prodfracBs}
\end{equation}
 where $N_{\Bs}(\mathrm{sl}) = 6222 \pm 144$, $N_{\Bs}(\mathrm h) = 5582 \pm 100$ are the
number of semileptonic and hadronic \Bs\ decays in the sample.  A
likelihood penalty of $0.5\left[
(A_{\mathrm{sl,h}} - A_{\mathrm{sl,h}}(\Bs))/\Delta A_{\mathrm{sl,h}}(\Bs)\right]^2$  is applied 
where $\Delta A_{\mathrm{sl,h}}(\Bs) = 0.020$ is the uncertainty. This uncertainty 
 includes the statistical uncertainty in the number of \Bs\
events and  the uncertainties in the relative reconstruction
efficiencies  and acceptances between the hadronic and semileptonic
data. A ratio has been chosen for the constraint as it is well behaved 
if either of the event yields ($N_{\mathrm h}$ and $N_{\mathrm{sl}}$) approaches zero. 

The fit results are summarized in Table~\ref{tab:tablecombined}
and the correlations between the fit parameters are given in
Table~\ref{tab:tablecorrelations}. The correlation of nearly one between
$N_{X}(\mathrm{sl})$ and $N_{X}(\mathrm{had})$ is a result of the
constraint on the event yields [Eq.~(\ref{eq:prodfrac})]. The local
statistical significance of the signal  is defined as \sig, where
${\cal{L}}_\mathrm{max}$ and ${\cal{L}}_0$ are likelihood values at the
best-fit signal yield and the signal yield fixed to zero obtained from a
binned maximum-likelihood fit. For the cone cut the  
  $p$-value of the fit to the data  with the cone cut 
 is $2.2 \times 10^{-14}$ and the local statistical significance is
7.6\,$\sigma$.
The $p$-value 
without the cone cut is  $8.2 \times 10^{-9}$ and the local statistical
significance is 5.8\,$\sigma$. 

\begin{table*}[hbt]
\def\arraystretch{1.25}
\caption{\label{tab:tablecombined} Results for the combined fit to the hadronic and semileptonic data sets (see Fig.~\ref{fig:fit_combined}).}
\begin{ruledtabular}
\begin{tabular}{ccc}
 & Cone cut & No cone cut \\
\hline
Fitted mass, MeV/$c^2$ 							& $5566.9^{+3.2}_{-3.1}\,\mbox{(stat)}^{+0.6}_{-1.2}\,\mbox{(syst)}$	& $5565.8^{+4.2}_{-4.0}\,\mbox{(stat)}^{+1.3}_{-2.0}\,\mbox{(syst)}$ 	\\
Fitted width, MeV/$c^2$ 						& $18.6^{+7.9}_{-6.1}\,\mbox{(stat)}^{+3.5}_{-3.8}\,\mbox{(syst)}$  	& $16.3^{+9.8}_{-7.6}\,\mbox{(stat)}^{+4.2}_{-6.5}\,\mbox{(syst)}$  	\\
Fitted number of hadronic signal events 		& $131^{+37}_{-33}\,\mbox{(stat)}^{+15}_{-14}\,\mbox{(syst)}$ 			& $99^{+40}_{-34}\,\mbox{(stat)}^{+18}_{-33}\,\mbox{(syst)}$ 	\\
Fitted number of semileptonic signal events 	& $147^{+42}_{-37}\,\mbox{(stat)}^{+17}_{-16}\,\mbox{(syst)}$ 			& $111.7^{+46}_{-39}\,\mbox{(stat)}^{+20}_{-38}\,\mbox{(syst)}$ 	\\
$\chi^{2}/$ndf 									& $94.7/(100-6)$  			& $54.2/(50-6)$  			\\
$p$-value										& $2.2 \times 10^{-14}$     & $1.9 \times 10^{-8}$      \\
Local significance 								& $7.6\, \sigma$  			& $5.6\, \sigma$  			\\
Significance with LEE							& $6.9\, \sigma$  			& $5.0\, \sigma$  			\\
Significance with LEE+systematics				& $6.7\, \sigma$  			& $4.7\, \sigma$  			\\
\end{tabular}
\end{ruledtabular}
\end{table*}

\begin{table*}[htb]
\def\arraystretch{1.25}
\caption{\label{tab:tablecorrelations} Correlations between the parameters  of the combined fit to the hadronic and semileptonic data sets (see Fig.~\ref{fig:fit_combined}). 
The yield in the semileptonic channel is $N_{X}(\mathrm{sl})$, the hadronic channel $N_{X}(\mathrm{h})$, 
while the fraction of background events is $f_{\rm sl,bgr}$ and $f_{\rm h,bgr}$ respectively. 
}
\begin{ruledtabular}
\begin{tabular}{ccccccc}
\multicolumn{7}{c}{ Cone Cut  }\\
\hline
		&  Mass		& 	Width  	& $N_{X}(\mathrm{sl})$	& $N_{X}(\mathrm{h})$	& $f_{\rm sl,bgr}$ & $f_{\rm h,bgr}$\\
\hline
Mass 							&    1   	&  0.22  	&   0.37  				&   0.37 				&  -0.06  	& -0.11 	\\
Width							&    0.22   &  1  		&   0.58  	&   0.59 	&  -0.16  	& -0.29 	\\
$N_{X}(\mathrm{sl})$			&    0.37   &  0.58  	&   1  		&   0.98 	&  -0.31  	& -0.44 	\\
$N_{X}(\mathrm{h})$ 			&    0.37   &  0.59  	&   0.98  	&   1 		&  -0.30  	& -0.45 	\\
$f_{\rm sl,bgr}$				&   -0.06   & -0.16  	&  -0.31  	&  -0.30 	&   1  		&  0.14 	\\
$f_{\rm h,bgr}$					&   -0.11   & -0.29  	&  -0.44  	&  -0.45 	&   0.14  	&  1 		\\
\hline
\multicolumn{7}{c}{ No Cone Cut  }\\
\hline
		&  Mass		& 	Width  	& $N_{X}(\mathrm{sl})$	& $N_{X}(\mathrm{h})$	& $f_{\rm sl,bgr}$ & $f_{\rm h,bgr}$\\
\hline
Mass 							&      1     &    0.38  &   0.49 &     0.49 &   -0.11 &   -0.17 \\ 
Width							&      0.38  &    1     & 	 0.64 &     0.64 &   -0.18 &   -0.31 \\ 
$N_{X}(\mathrm{sl})$			&      0.49  &    0.64  &   1    &  	0.99 &   -0.33 &   -0.45 \\ 
$N_{X}(\mathrm{h})$ 			&      0.49  &    0.64  &   0.99 &     1    & 	 -0.33 &   -0.46 \\ 
$f_{\rm sl,bgr}$				&     -0.11  &   -0.18  &  -0.33 &    -0.33 &    1    & 	0.15 \\
$f_{\rm h,bgr}$ 				&     -0.17  &   -0.31  &  -0.45 &    -0.46 &    0.15 &    1 \\
\end{tabular}
\end{ruledtabular}
\end{table*}

\subsection{Systematic uncertainties}

The systematic uncertainties of the combined fit are given in
Table~\ref{tab:table11}. The  uncertainty  on  (i) the
background shape descriptions is evaluated  by
using the alternative paramaterizations of the background, Eqs. (2), (3)
and the smoothed MC histogram independently for the semileptonic and the
hadronic channels (16 different fits) and finding the maximal deviations
from the nominal fit. 

The effect of (ii) the MC weighting  for the
semileptonic background is estimated by creating 1000 background samples
where the weights have been randomly varied based on the uncertainties
in the weighting procedure and measuring the standard deviation and bias
of the measured values.  

The (iii) MC component of the background for the
hadronic sample is made up of a mixture of two independent MC samples
with different production properties (see Ref.~\cite{D0:2016mwd}) and
the systematic uncertainties due to this are found by varying the composition of this
mixture and
measuring the standard deviation and bias of the measured values. 
The (iv) size of the hadronic sidebands is varied using the maximal 
deviations from the nominal fit to estimate the systematic uncertainty.

The
systematic uncertainty due to the (v) fraction of MC and SS data in the semileptonic sample,
(vi) the MC and sideband data in
the case of the hadronic, is varied independently between the two
samples  measuring the standard deviation and bias of the measured
values. Since the background model for the semileptonic sample without the
cone cut only uses the MC background simulation this uncertainty (v) does not apply. 

All of the uncertainties due to the
modeling of the background are assumed to be independent for the
hadronic and semileptonic data samples.

The remaining uncertainties are measured by finding the maximal
deviations from the nominal fit for (vii)  varying the energy scale in the
semileptonic and hadronic MC data samples by $\pm$1\,MeV/$c^2$ in both
samples simultaneously; (viii)  varying the nominal  mass resolution of
3.85\,MeV/$c^2$ for the D0 detector  by $\pm$1\,MeV/$c^2$ and 
$+2$\,MeV/$c^2$ in both the hadronic and semileptonic data samples simultaneously; 
(ix) varying the resolution of the $X^\pm(5568)$ peak in the
semileptonic channel either by $\pm$1\,MeV/$c^2$ around the mean value given by
Eq.~(\ref{eq:mass_resolution}) or by using a constant resolution
of 11.1\,MeV/$c^2$ for the semileptonic data while the mass resolution in the
hadronic channel remains at 3.85\,MeV/$c^2$; (x) using a P-wave
relativistic Breit-Wigner function for both data sets; (xi) setting the shift
of the fitted mass peak in the semileptonic data with respect to the
hadronic data  due to the missing neutrino to  $\pm$1\,MeV/$c^2$; and
(xii) varying the   constraint on the relative number of signal events in
hadronic and semileptonic channels (Eq.~(\ref{eq:prodfrac})) between 0.034
and 0.074. The correlation of each of the sources of systematic
uncertainty between the hadronic and semileptonic data sets is
indicated in Table~\ref{tab:table11}. The uncertainties are added in
quadrature separately for positive and negative values to obtain the
total systematic uncertainties for each measured parameter.
The results including systematic uncertainties are given in Table~\ref{tab:tablecombined}.

\begin{table*}[htbp]
\def\arraystretch{1.25}
\caption{\label{tab:table11} Systematic uncertainties  of the combined fit for the 
$X^\pm(5568)$ state mass,  width and the event yields. 
Each uncertainty is either correlated or uncorrelated between the hadronic and semileptonic data sets. }
\begin{ruledtabular}
\begin{tabular}{lccccc}
Source & Sample & Mass, MeV/$c^2$ & Width, MeV/$c^2$  & \multicolumn{2}{c}{Event yields, events}  \\
\multicolumn{4}{c}{} & hadronic &  semileptonic\\
\hline 
\multicolumn{6}{c}{ Cone Cut  }\\
\hline
(i) Background shape description 		& Both		 	& $+0.3$ \ ;\ $-0.6$ & $+1.9$ \ ;\ $-0.0$ & $+0.0$ \ ;\ $-6.6$ & $+0.0$ \ ;\ $-7.8$ \\
(ii) SL background reweighting 			& Semileptonic 	& $+0.1$ \ ;\ $-0.2$ & $+0.2$ \ ;\ $-0.2$ & $+2.5$ \ ;\ $-3.3$ & $+2.9$ \ ;\ $-3.9$ \\
(iii) Hadronic MC samples				& Hadronic		& $+0.3$ \ ;\ $-0.2$ & $+1.2$ \ ;\ $-0.4$ & $+7.0$ \ ;\ $-2.5$ & $+7.8$ \ ;\ $-2.8$ \\
(iv) Hadronic sidebands					& Hadronic  	& $+0.1$ \ ;\ $-0.1$ & $+0.5$ \ ;\ $-1.3$ & $+2.3$ \ ;\ $-9.3$ & $+2.5$ \ ;\ $-10.2$ \\
(v) SL MC/Data ratio 	 				& Semileptonic  & $+0.0$ \ ;\ $-0.1$ & $+0.1$ \ ;\ $-0.1$ & $+1.0$ \ ;\ $-1.2$ & $+1.1$ \ ;\ $-1.4$ \\
(vi) Hadronic MC simulation/data ratio				& Hadronic      & $+0.0$ \ ;\ $-0.0$ & $+0.2$ \ ;\ $-0.2$ & $+1.0$ \ ;\ $-1.1$ & $+1.1$ \ ;\ $-1.2$ \\
(vii) $B_s^0$ mass scale, MC simulation and data 	& Both		  	& $+0.2$ \ ;\ $-0.2$ & $+0.8$ \ ;\ $-0.8$ & $+3.7$ \ ;\ $-4.3$ & $+4.1$ \ ;\ $-4.7$ \\
(viii) Detector resolution 				& Both		 	& $+0.1$ \ ;\ $-0.3$ & $+1.3$ \ ;\ $-3.4$ & $+1.4$ \ ;\ $-3.8$ & $+1.6$ \ ;\ $-4.2$ \\
(ix) Missing neutrino effect  			& Semileptonic 	& $+0.1$ \ ;\ $-0.1$ & $+0.1$ \ ;\ $-0.0$ & $+0.5$ \ ;\ $-0.1$ & $+0.0$ \ ;\ $-0.4$ \\
(x) {\it P}-wave Breit-Wigner  			& Both		 	& $+0.0$ \ ;\ $-0.0$ & $+2.1$ \ ;\ $-0.0$ & $+11.7$ \ ;\ $-0.0$ & $+13.0$ \ ;\ $-0.0$ \\
(xi) Mass offset						& Both		    & $+0.3$ \ ;\ $-0.3$ & $+0.1$ \ ;\ $-0.0$ & $+0.2$ \ ;\ $-0.4$ & $+0.3$ \ ;\ $-0.4$ \\
(xii) Production fraction 				& Both		    & $+0.0$ \ ;\ $-0.0$ & $+0.1$ \ ;\ $-0.1$ & $+1.4$ \ ;\ $-1.6$ & $+4.2$ \ ;\ $-4.2$ \\
\hline
Total 									&			    & $+0.6$ \ ;\ $-1.2$ & $+3.5$ \ ;\ $-3.8$ & $+14.7$ \ ;\ $-13.6$ & $+16.9$ \ ;\ $-15.8$ \\
\hline
\multicolumn{6}{c}{No Cone Cut}\\
\hline
(i) Background shape description 		& Both		 	& $+1.1$ \ ;\ $-1.9$ & $+1.4$ \ ;\ $-5.1$ & $+7.6$ \ ;\ $-32.8$ & $+8.4$ \ ;\ $-37.1$ \\
(ii) SL background reweighting 			& Semileptonic 	& $+0.1$ \ ;\ $-0.0$ & $+0.1$ \ ;\ $-0.3$ & $+1.8$ \ ;\ $-1.1$ & $+2.0$ \ ;\ $-1.4$ \\
(iii) Hadronic MC samples				& Hadronic		& $+0.3$ \ldots\ \ ;\ $-0.0$ & $+1.1$ \ ;\ $-0.0$ & $+7.2$ \ ;\ $-0.0$ & $+7.9$ \ ;\ $-0.0$ \\
(iv) Hadronic Sidebands					& Hadronic  		& $+0.3$ \ ;\ $-0.1$ & $+0.2$ \ ;\ $-0.6$ & $+4.5$ \ ;\ $-3.7$ & $+4.9$ \ ;\ $-4.2$ \\
(v) SL MC simulation/data ratio					& Not applicable  	& \ldots\ \ ;\ \ldots\ & \ldots\ \ ;\ \ldots\ & \ldots\ \ ;\ \ldots\ & \ldots\ \ ;\ \ldots\ \\
(v) Hadronic MC simulation/data ratio				& Hadronic      	& $+0.1$ \ ;\ $-0.0$ & $+0.5$ \ ;\ $-0.0$ & $+7.4$ \ ;\ $-0.1$ & $+8.1$ \ ;\ $-0.2$ \\
(vii) $B_s^0$ mass scale, MC simulation and data 	& Both		  	& $+0.1$ \ ;\ $-0.1$ & $+0.9$ \ ;\ $-0.2$ & $+5.1$ \ ;\ $-0.0$ & $+5.6$ \ ;\ $-0.0$ \\
(viii) Detector resolution 				& Both		 	& $+0.1$ \ ;\ $-0.2$ & $+1.6$ \ ;\ $-3.9$ & $+1.5$ \ ;\ $-3.5$ & $+1.6$ \ ;\ $-4.0$ \\
(ix) Missing neutrino effect  			& Semileptonic 	& $+0.2$ \ ;\ $-0.1$ & $+0.1$ \ ;\ $-0.1$ & $+0.4$ \ ;\ $-0.0$ & $+0.1$ \ ;\ $-0.3$ \\
(x) {\it P}-wave Breit-Wigner  			& Both		 	& $+0.0$ \ ;\ $-0.6$ & $+3.3$ \ ;\ $-0.0$ & $+10.7$ \ ;\ $-0.0$ & $+11.8$ \ ;\ $-0.0$ \\
(xi) Mass offset						& Both		    & $+0.4$ \ ;\ $-0.4$ & $+0.2$ \ ;\ $-0.2$ & $+0.0$ \ ;\ $-0.0$ & $+0.0$ \ ;\ $-0.1$ \\
(xii) Production fraction 				& Both		    & $+0.0$ \ ;\ $-0.0$ & $+0.1$ \ ;\ $-0.1$ & $+0.8$ \ ;\ $-0.8$ & $+3.5$ \ ;\ $-3.6$ \\
\hline
Total 									&			    & $+1.3$ \ ;\ $-2.0$ & $+4.2$ \ ;\ $-6.5$ & $+18.2$ \ ;\ $-33.2$ & $+20.3$ \ ;\ $-37.8$ \\
\end{tabular}
\end{ruledtabular}
\end{table*}

\subsection{Significance}

The look-elsewhere effect (LEE) is determined using the approach proposed in
Ref.~\cite{Gross:2010qma}. We have generated 250,000 simulated
background distributions with no signal, both with and without the cone cut. These
distributions are fit using the same procedure as the data. The mass
parameter of the relativistic Breit-Wigner is constrained  to be between 
5506 to 5675\,MeV/$c^2$ (the sum of the mass of the $B^0_d$ and $K^\pm$)
with a starting value of $m_X = 5600$\,MeV/$c^2$. The  width of
the signal is allowed to vary between 0.1 and 60\,MeV/$c^2$ with a
starting value of $\Gamma_X = 21$\,MeV/$c^2$. The maximum local
statistical significance for each distribution is calculated. The resulting
distribution of the local significance is fitted with the function 
\begin{equation}
f_{\mathrm{loc}} = N_{\mathrm{trials}} \left[ \chi^2(2) + P_1 \chi^2(3) \right],
\label{eq:floc}
\end{equation} 
where $N_{\mathrm{trials}}$ is the number of generated distributions, $P_1$ is
a free parameter and $\chi^2(n)$ is the $\chi^2$ cumulative distribution
function for $n$ degrees of freedom. We have used $n=2$ and 3 as we are
fitting two spectra simultaneously.  The resulting function is
integrated above the measured local significance to determine the global
significance (Table~\ref{tab:tablecombined}). The significance, not 
including the systematic uncertainty, of the observed
signal accounting for the LEE and with the cone cut applied is
6.9\,$\sigma$ ($p$-value = $4.1 \times 10^{-12}$). The significance of
the signal without the cone cut is 5.0\,$\sigma$ ($p$-value = $4.1
\times 10^{-7}$). The effect of choosing the function in Eq.~(\ref{eq:floc}) is studied by modifying it
to $f_{\mathrm{loc}} = N_{\mathrm{trials}} \left[ \chi^2(2) + P_1
\chi^2(4) \right]$ and $f_{\mathrm{loc}} = N_{\mathrm{trials}} \left[
\chi^2(2) + P_1 \chi^2(3) + P_2 \chi^2(4) \right]$ with no significant
change to the significance being observed.
The  look-elsewhere effect on the signal significance
is checked with a method described in Ref.~\cite{Gross:2010qma} that 
relates the tail probability with the number of 
``upcrossing'' at a small reference level. Five hundred simulated background 
spectra are generated. Each of these 500 distributions is fitted with the background 
plus signal function with different initial masses from 5506 to 5675\,MeV/$c^2$ in 5\,MeV/$c^2$ 
steps along with a background-only fit. The significance  is plotted for each of the mass points 
and the number of upcrossings (each time the significance crosses a small reference value) is measured.  
The mean number of upcrossings for a reference level of 0.5 is determined and the global 
significance is calculated. The resulting significance is consistent with the method described above.

The systematic uncertainties are treated as nuisance parameters to
construct a prior predictive model~\cite{Olive:2016xmw,Giunti:1998xv} of
our test statistic. When the systematic uncertainties are included, the
significance of the observed signal with the cone cut applied for the
combined fit is reduced to 6.7\,$\sigma$ ($p$-value = $1.5 \times
10^{-11}$) and the significance of the signal without the cone cut is
4.7\,$\sigma$ ($p$-value = $2.0 \times 10^{-6}$).

\subsection{Closure tests}

\begin{figure*}[htbp]
  \subfloat[Semileptonic Sample]{\includegraphics[width=0.49\linewidth]{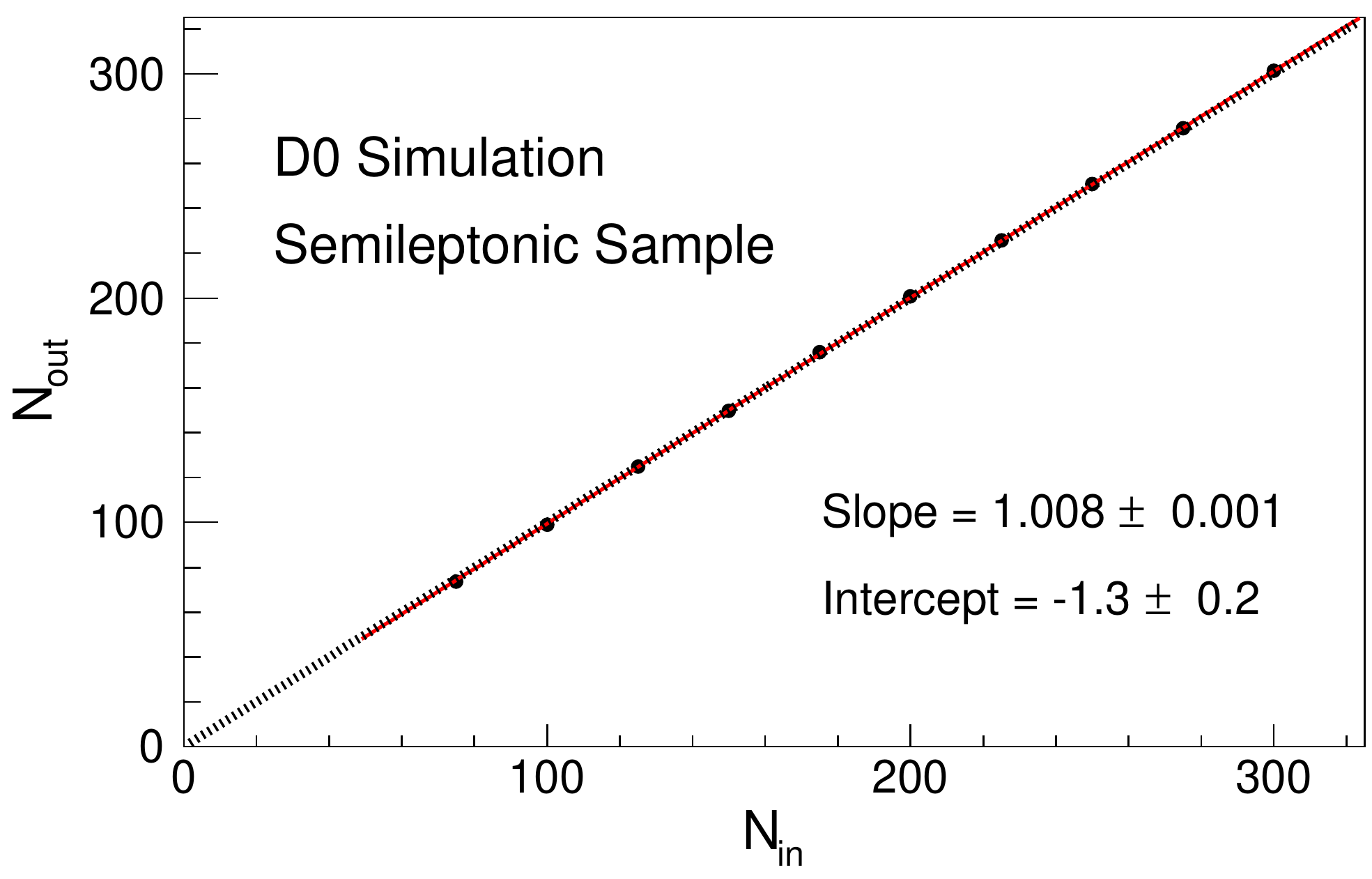}}
  \subfloat[Hadronic Sample]{\includegraphics[width=0.49\linewidth]{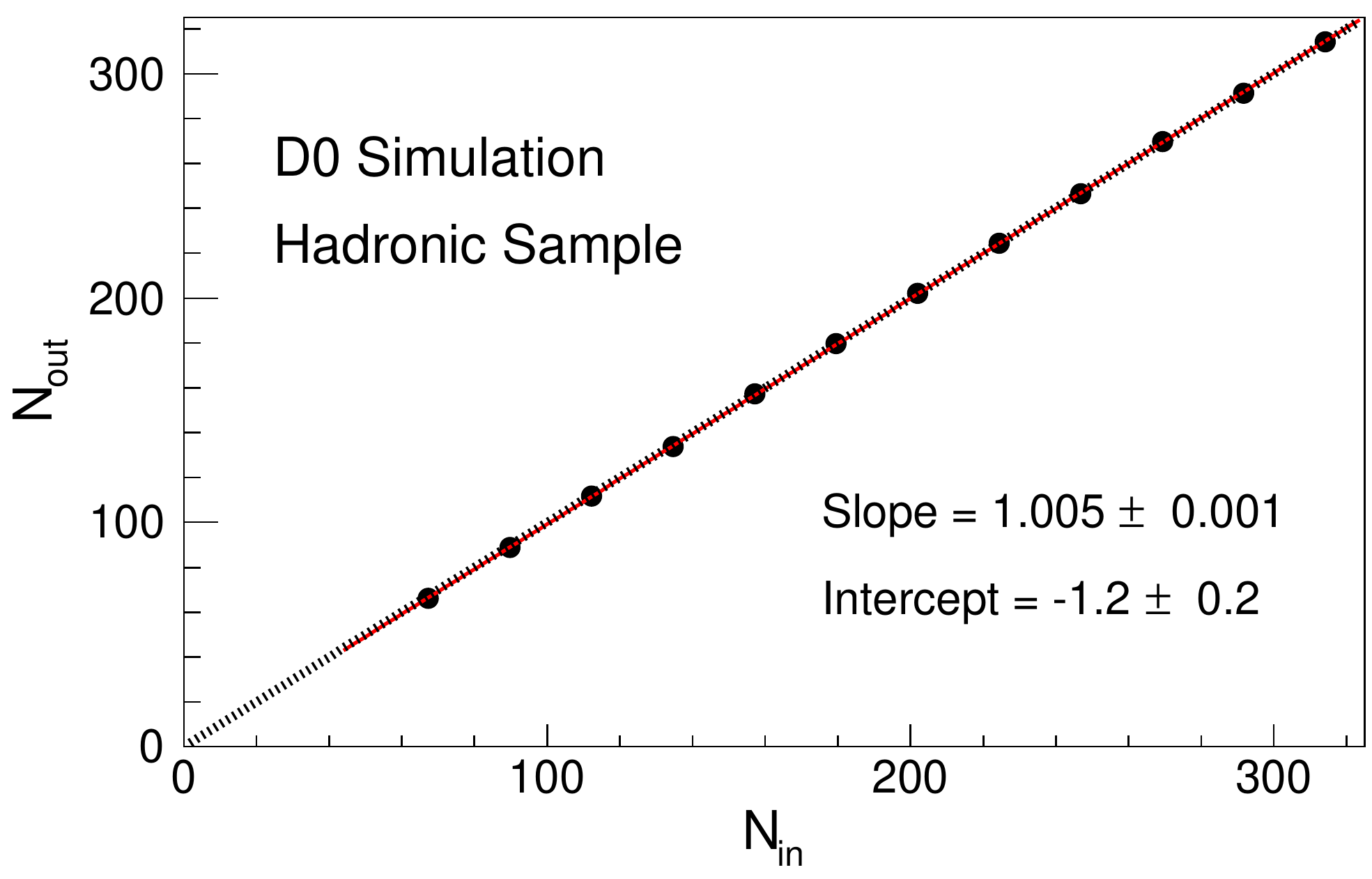}}
\caption{\label{fig:si_plot} Results of the toy MC tests of the combined sample fitting
procedure (black circles) used in the analysis with   the cone cut.  The number of fitted signal events
 are plotted vs    fitted number of injected
signal events for the (a) semileptonic and (b) hadronic samples. The dotted line shows
$N_\mathrm{in}=N_\mathrm{out}$ and the red line shows the fit to a line.}
\end{figure*}

To test the sensitivity and accuracy of the fitting procedure for the combined signal 
extraction we repeat the closure tests carried out in Sec.~\ref{sec:closure1}
with the following modifications. 
The size of the associated hadronic signal is set using Eqs.~(11) and (12).
The appropriate detector
resolution is used, Eq.~(\ref{eq:mass_resolution}) for the semileptonic sample and 3.85\,MeV/$c^2$ 
for the hadronic sample.
For each trial the fitting procedure is
performed to obtain the  mass and width and the number of semileptonic and hadronic signal events.
The results of each set of trials is fitted with a Gaussian to determine
the mean and the uncertainty in the number of signal events, the mass
and the width (see Table~\ref{tab:tabletoyMC}). The number of fitted
signal events vs. the number of injected signal events for the semileptonic and hadronic samples is plotted in
Fig.~\ref{fig:si_plot}. These results show excellent agreement between the input and fit parameters.

\begin{table*}[hbtp]
\def\arraystretch{1.25}
\caption{\label{tab:tabletoyMC} Mean values and uncertainties for fitted number of events, 
 mass and width  from
 Gaussian fits to corresponding distributions from 10,000 trials with the cone cut.
 Also given is the expected statistical uncertainties on the fitted number of events, $\Delta$($N_{\mathrm{fit}}$),
 and the expected uncertainties on the measurement of the width, $\Delta(\Gamma_X)$\,MeV/$c^2$. 
   A range
 of signals  with 75, 100, 125, 150, 175 and 200 signal events, mass $m_x
 = 5568.3$\,MeV$/c^2$ and width $\Gamma_X = 21.9$\,MeV$/c^2$ have been
 simulated. Background parametrization Eq.~(\ref{eq:ap_bkg}) is used.}
\begin{ruledtabular}
\begin{tabular}{ccccccccc}
\multicolumn{3}{c}{Semileptonic channel} & \multicolumn{3}{c}{Hadronic channel} & $m_X$ & $\Gamma_X$ & $\Delta(\Gamma_X)$\\
$N_{\mathrm{in}}(\mathrm{sl})$ & $N_{\mathrm{fit}}(\mathrm{sl})$ & $\Delta$($N_{\mathrm{fit}}(\mathrm{sl})$) &
$N_{\mathrm{in}}(\mathrm{h})$ & $N_{\mathrm{fit}}(\mathrm{h})$ & $\Delta$($N_{\mathrm{fit}}(\mathrm{h})$) &
MeV/$c^2$ & MeV/$c^2$  & MeV/$c^2$ \\
\hline
75	& $73.8	 \pm	0.3$ &	25.7 & 67.3	 &	$66.0	\pm 0.2$ & 23.0 &	$5569.0 \pm	0.076$	& 19.3 & 10.9	 \\
100	& $99.1	 \pm	0.3$ &	26.3 & 89.8  &	$88.7	\pm 0.2$ & 23.6 &	$5568.4 \pm	0.042$	& 20.8 & 9.2 	\\
125	& $124.9 \pm	0.3$ &	26.8 & 112.2 &	$111.7	\pm 0.2$ & 24.0 &	$5568.4 \pm	0.032$	& 21.5 & 7.8 	\\
150	& $149.6 \pm	0.3$ &	26.5 & 134.6 &	$133.8	\pm 0.2$ & 23.6 &	$5568.4 \pm	0.027$	& 21.9 & 6.8 	\\
175	& $175.9 \pm	0.3$ &	27.2 & 157.1 &	$157.3	\pm 0.2$ & 24.3 &	$5568.4 \pm	0.023$	& 22.3 & 6.0 	\\
200	& $200.8 \pm	0.3$ &	27.2 & 179.5 &	$179.6	\pm 0.2$ & 24.2 &	$5568.4 \pm	0.021$	& 22.4 & 5.4 	\\
\end{tabular}
\end{ruledtabular}
\end{table*}

\subsection{Cross-checks}

 To test the stability of the results, alternative choices are made regarding the fit parameters 
 (see Table~\ref{tab:tablecombinedcrosschecks}). 
 
When no constraint is placed  on the ratio of the event yields in 
the hadronic and semileptonic channels, Eq.~(\ref{eq:prodfrac}), the results  
are entirely consistent with the fit with the constraint. 
 
 We have also carried out a fit in which two of the systematic effects are treated as nuisance parameters 
 in the fit. We allow a mass shift, $\Delta m$, between the hadronic and semileptonic data with a likelihood penalty of 
 $0.5(\Delta m/1 \,\mathrm{MeV}/c^2)^2$. We also allow the overall resolution of the semileptonic signal to vary by $\Delta \sigma_{\mathrm{SL}}$
 with a likelihood penalty of 
 $0.5(\Delta \sigma_{\mathrm{SL}}/1 \,\mathrm{MeV}/c^2)^2$. The resultant fit produces a mass,  width and event yields that are consistent  
 with the default fit and shifts of $\Delta m = (0.0 \pm 1.4)$\,MeV/$c^2$ and  $\Delta \sigma_{\mathrm{SL}} = (-0.1 \pm 1.4)$\,MeV/$c^2$.

 The significance of a nonzero  width is determined by fitting the data with the  width set to zero and 
 comparing it with the fit with no constraint on the width (Table~\ref{tab:tablecombinedcrosschecks}). 
 Using the data with the cone cut the $p$-value of the  width being consistent  with zero is 
 $5.4 \times 10^{-6}$, and the  statistical significance is 4.5\,$\sigma$. 
 The significance without the cone cut is 
  3.3\,$\sigma$ ($p$-value = $1.1 \times 10^{-3}$).
 
\begin{table*}[htb]
\def\arraystretch{1.25}
\caption{\label{tab:tablecombinedcrosschecks} Various cross checks for the combined fit of the hadronic and semileptonic data sets.}
\begin{ruledtabular}
\begin{tabular}{ccccc}
 & Default fit  & No production constraint & Nuisance parameter & Zero width\\
\hline
\multicolumn{5}{c}{Cone Cut}\\
\hline
Fitted mass, MeV/$c^2$ 							& $5566.9^{+3.2}_{-3.1}$  	& $5566.8^{+3.2}_{-3.1}$	& $5567.4^{+3.2}_{-3.4}$	& $5569.9^{+1.3}_{-1.3}$\\
Fitted width, MeV/$c^2$ 						& $18.6^{+7.9}_{-6.1}$  	& $18.3^{+8.0}_{-6.2}$  	& $21.7^{+7.3}_{-5.5}$  	& $0$  \\
Fitted number of hadronic signal events 		& $131^{+37}_{-33}$ 		& $127^{+34}_{-29}$ 		& $134^{+37}_{-33}$ 		& $60^{+17}_{-16}$ \\
Fitted number of semileptonic signal events 	& $147^{+42}_{-37}$ 		& $159^{+66}_{-59}$ 		& $151^{+41}_{-37}$ 		& $68^{+19}_{-18}$ \\
$\chi^{2}/$ndf 									& $94.7/(100-6)$  			& $94.5/(100-6)$  			& $94.8/(100-8)$  			& $115.4/(100-7)$  		\\
$p$-value										& $2.2 \times 10^{-14}$   	& $2.0 \times 10^{-14}$     & $2.4 \times 10^{-14}$     & $8.5 \times 10^{-10}$    \\
Local significance 								& $7.6\, \sigma$  			& $7.7\, \sigma$  			& $7.6\, \sigma$  			& $6.1\, \sigma$  			\\
\hline
\multicolumn{5}{c}{No cone cut}\\
\hline
Fitted mass, MeV/$c^2$ 							& $5565.8^{+4.2}_{-4.0}$ 	& $5565.8^{+4.1}_{-3.9}$ 	& $5566.3^{+4.4}_{-4.6}$ 	& $5569.7^{+1.6}_{-1.9}$		\\
Fitted width, MeV/$c^2$ 						& $16.3^{+9.8}_{-7.6}$  	& $15.0^{+9.6}_{-7.8}$  	& $20.0^{+9.1}_{-9.4}$  	& $0$  							\\
Fitted number of hadronic signal events 		& $99^{+40}_{-34}$ 			& $84^{+43}_{-35}$ 			& $103^{+40}_{-37}$ 		& $48^{+17}_{-16}$ 		\\
Fitted number of semileptonic signal events 	& $112^{+46}_{-39}$ 		& $151^{+72}_{-61}$ 		& $115^{+45}_{-42}$ 		& $54^{+20}_{-18}$ 		\\
$\chi^{2}/$ndf 									& $54.2/(50-6)$  			& $52.5/(50-6)$  			& $54.8/(50-8)$  			& $101.3/(50-7)$  				\\
$p$-value										& $1.9 \times 10^{-8}$      & $8.2 \times 10^{-9}$      & $2.7 \times 10^{-8}$      & $5.1 \times 10^{-6}$    		\\
Local significance 								& $5.6\, \sigma$  			& $5.8\, \sigma$  			& $5.6\, \sigma$  			& $4.6\, \sigma$  				\\
\end{tabular}
\end{ruledtabular}
\end{table*}

\section{Conclusions}

We have presented the results of a search for the $X^\pm(5568) \to \Bs\pi^\pm$ with
semileptonic decays of the $B_s^0$ meson.
The  $X^\pm(5568) \to B_s^0 \pi^\pm$ state reported in the case that 
$\Bs \to J/\psi\phi$~\cite{D0:2016mwd} is confirmed for the case that 
$B_s^0 \rightarrow \mu^\mp D_s^\pm \, \mathrm{X}$, \Dsdecay .
The analyses of the hadronic and semileptonic  data give similar measurements 
of the mass, width and  production ratio of $X^\pm(5568)$ to a \Bs\ meson.  
The mass and   width of this state are measured using a combined fit of both data sets  with the cone cut, 
yielding $m = 5566.9 ^{+3.2}_{-3.1} \thinspace  {\rm (stat)} ^{+0.6}_{-1.2}  {\rm \thinspace (syst)}$\,MeV/$c^2$,
$\Gamma = 18.6 ^{+7.9}_{-6.1}  {\rm \thinspace (stat)}   ^{+3.5}_{-3.8} {\rm \thinspace (syst)} $\,MeV/$c^2$. 
The p-value for the null signal hypothesis to represent the data is $1.5 \times 10^{-11}$  (6.7$\,\sigma$).

\section{Acknowledgements}

%

This document was prepared by the D0 collaboration using the resources of the Fermi National Accelerator Laboratory (Fermilab),
a U.S. Department of Energy, Office of Science, HEP User Facility. Fermilab is managed by Fermi Research Alliance, LLC (FRA),
acting under Contract No. DE-AC02-07CH11359.

We thank the staffs at Fermilab and collaborating institutions,
and acknowledge support from the
Department of Energy and National Science Foundation (United States of America);
Alternative Energies and Atomic Energy Commission and
National Center for Scientific Research/National Institute of Nuclear and Particle Physics  (France);
Ministry of Education and Science of the Russian Federation, 
National Research Center ``Kurchatov Institute" of the Russian Federation, and 
Russian Foundation for Basic Research  (Russia);
National Council for the Development of Science and Technology and
Carlos Chagas Filho Foundation for the Support of Research in the State of Rio de Janeiro (Brazil);
Department of Atomic Energy and Department of Science and Technology (India);
Administrative Department of Science, Technology and Innovation (Colombia);
National Council of Science and Technology (Mexico);
National Research Foundation of Korea (Korea);
Foundation for Fundamental Research on Matter (The Netherlands);
Science and Technology Facilities Council and The Royal Society (United Kingdom);
Ministry of Education, Youth and Sports (Czech Republic);
Bundesministerium f\"{u}r Bildung und Forschung (Federal Ministry of Education and Research) and 
Deutsche Forschungsgemeinschaft (German Research Foundation) (Germany);
Science Foundation Ireland (Ireland);
Swedish Research Council (Sweden);
China Academy of Sciences and National Natural Science Foundation of China (China);
and
Ministry of Education and Science of Ukraine (Ukraine).
%



\begin{thebibliography}{33}%
\makeatletter
\providecommand \@ifxundefined [1]{%
 \@ifx{#1\undefined}
}%
\providecommand \@ifnum [1]{%
 \ifnum #1\expandafter \@firstoftwo
 \else \expandafter \@secondoftwo
 \fi
}%
\providecommand \@ifx [1]{%
 \ifx #1\expandafter \@firstoftwo
 \else \expandafter \@secondoftwo
 \fi
}%
\providecommand \natexlab [1]{#1}%
\providecommand \enquote  [1]{``#1''}%
\providecommand \bibnamefont  [1]{#1}%
\providecommand \bibfnamefont [1]{#1}%
\providecommand \citenamefont [1]{#1}%
\providecommand \href@noop [0]{\@secondoftwo}%
\providecommand \href [0]{\begingroup \@sanitize@url \@href}%
\providecommand \@href[1]{\@@startlink{#1}\@@href}%
\providecommand \@@href[1]{\endgroup#1\@@endlink}%
\providecommand \@sanitize@url [0]{\catcode `\\12\catcode `\$12\catcode
  `\&12\catcode `\#12\catcode `\^12\catcode `\_12\catcode `\%12\relax}%
\providecommand \@@startlink[1]{}%
\providecommand \@@endlink[0]{}%
\providecommand \url  [0]{\begingroup\@sanitize@url \@url }%
\providecommand \@url [1]{\endgroup\@href {#1}{\urlprefix }}%
\providecommand \urlprefix  [0]{URL }%
\providecommand \Eprint [0]{\href }%
\providecommand \doibase [0]{http://dx.doi.org/}%
\providecommand \selectlanguage [0]{\@gobble}%
\providecommand \bibinfo  [0]{\@secondoftwo}%
\providecommand \bibfield  [0]{\@secondoftwo}%
\providecommand \translation [1]{[#1]}%
\providecommand \BibitemOpen [0]{}%
\providecommand \bibitemStop [0]{}%
\providecommand \bibitemNoStop [0]{.\EOS\space}%
\providecommand \EOS [0]{\spacefactor3000\relax}%
\providecommand \BibitemShut  [1]{\csname bibitem#1\endcsname}%
\let\auto@bib@innerbib\@empty
\bibitem [{\citenamefont {Gell-Mann}(1964)}]{GellMann:1964nj}%
  \BibitemOpen
  \bibfield  {author} {\bibinfo {author} {\bibfnamefont {M.}~\bibnamefont
  {Gell-Mann}},\ }\bibfield  {title} {\enquote {\bibinfo {title} {{A Schematic
  Model of Baryons and Mesons}},}\ }\href {\doibase
  10.1016/S0031-9163(64)92001-3} {\bibfield  {journal} {\bibinfo  {journal}
  {Phys. Lett.}\ }\textbf {\bibinfo {volume} {8}},\ \bibinfo {pages} {214}
  (\bibinfo {year} {1964})}\BibitemShut {NoStop}%
\bibitem [{\citenamefont {Zweig}(1964)}]{Zweig:1981pd}%
  \BibitemOpen
  \bibfield  {author} {\bibinfo {author} {\bibfnamefont {G.}~\bibnamefont
  {Zweig}},\ }\bibfield  {title} {\enquote {\bibinfo {title} {{An SU(3) model
  for strong interaction symmetry and its breaking. Version 1}},}\ }\href@noop
  {} {\bibfield  {journal} {\bibinfo  {journal} {Report No. CERN-TH-401}\ } (\bibinfo
  {year} {1964})}\BibitemShut {NoStop}%
\bibitem [{\citenamefont {Maiani}\ \emph {et~al.}(2004)\citenamefont {Maiani},
  \citenamefont {Piccinini}, \citenamefont {Polosa},\ and\ \citenamefont
  {Riquer}}]{Maiani:2004uc}%
  \BibitemOpen
  \bibfield  {author} {\bibinfo {author} {\bibfnamefont {L.}~\bibnamefont
  {Maiani}}, \bibinfo {author} {\bibfnamefont {F.}~\bibnamefont {Piccinini}},
  \bibinfo {author} {\bibfnamefont {A.~D.}\ \bibnamefont {Polosa}}, \ and\
  \bibinfo {author} {\bibfnamefont {V.}~\bibnamefont {Riquer}},\ }\bibfield
  {title} {\enquote {\bibinfo {title} {{A New look at scalar mesons}},}\ }\href
  {\doibase 10.1103/PhysRevLett.93.212002} {\bibfield  {journal} {\bibinfo
  {journal} {Phys. Rev. Lett.}\ }\textbf {\bibinfo {volume} {93}},\ \bibinfo
  {pages} {212002} (\bibinfo {year} {2004})}\BibitemShut {NoStop}%
\bibitem [{\citenamefont {Esposito}\ \emph {et~al.}(2016)\citenamefont
  {Esposito}, \citenamefont {Pilloni},\ and\ \citenamefont
  {Polosa}}]{Esposito:2016noz}%
  \BibitemOpen
  \bibfield  {author} {\bibinfo {author} {\bibfnamefont {A.}~\bibnamefont
  {Esposito}}, \bibinfo {author} {\bibfnamefont {A.}~\bibnamefont {Pilloni}}, \
  and\ \bibinfo {author} {\bibfnamefont {A.~D.}\ \bibnamefont {Polosa}},\
  }\bibfield  {title} {\enquote {\bibinfo {title} {{Multiquark Resonances}},}\
  }\href {\doibase 10.1016/j.physrep.2016.11.002} {\bibfield  {journal}
  {\bibinfo  {journal} {Phys. Rep.}\ }\textbf {\bibinfo {volume} {668}},\
  \bibinfo {pages} {1} (\bibinfo {year} {2017})}\BibitemShut {NoStop}%
\bibitem [{\citenamefont {Chen}\ \emph {et~al.}(2016)\citenamefont {Chen},
  \citenamefont {Chen}, \citenamefont {Liu},\ and\ \citenamefont
  {Zhu}}]{Chen:2016qju}%
  \BibitemOpen
  \bibfield  {author} {\bibinfo {author} {\bibfnamefont {H.}~\bibnamefont
  {Chen}}, \bibinfo {author} {\bibfnamefont {W.}~\bibnamefont {Chen}}, \bibinfo
  {author} {\bibfnamefont {X.}~\bibnamefont {Liu}}, \ and\ \bibinfo {author}
  {\bibfnamefont {S.}~\bibnamefont {Zhu}},\ }\bibfield  {title} {\enquote
  {\bibinfo {title} {{The hidden-charm pentaquark and tetraquark states}},}\
  }\href {\doibase 10.1016/j.physrep.2016.05.004} {\bibfield  {journal}
  {\bibinfo  {journal} {Phys. Rept.}\ }\textbf {\bibinfo {volume} {639}},\
  \bibinfo {pages} {1--121} (\bibinfo {year} {2016})}\BibitemShut {NoStop}%
\bibitem [{\citenamefont {Olsen}\ \emph {et~al.}(2018)\citenamefont {Olsen},
  \citenamefont {Skwarnicki},\ and\ \citenamefont {Zieminska}}]{Olsen:2017bmm}%
  \BibitemOpen
  \bibfield  {author} {\bibinfo {author} {\bibfnamefont {S.~L.}\ \bibnamefont
  {Olsen}}, \bibinfo {author} {\bibfnamefont {T.}~\bibnamefont {Skwarnicki}}, \
  and\ \bibinfo {author} {\bibfnamefont {D.}~\bibnamefont {Zieminska}},\
  }\bibfield  {title} {\enquote {\bibinfo {title} {{Nonstandard heavy mesons
  and baryons: Experimental evidence}},}\ }\href {\doibase
  10.1103/RevModPhys.90.015003} {\bibfield  {journal} {\bibinfo  {journal}
  {Rev. Mod. Phys.}\ }\textbf {\bibinfo {volume} {90}},\ \bibinfo {pages}
  {015003} (\bibinfo {year} {2018})}\BibitemShut {NoStop}%
\bibitem [{\citenamefont {Choi}\ \emph {et~al.}(2003)\citenamefont {Choi} \emph
  {et~al.}}]{Choi:2003ue}%
  \BibitemOpen
  \bibfield  {author} {\bibinfo {author} {\bibfnamefont {S.~K.}\ \bibnamefont
  {Choi}} \emph {et~al.} (\bibinfo {collaboration} {Belle}),\ }\bibfield
  {title} {\enquote {\bibinfo {title} {{Observation of a narrow charmonium -
  like state in exclusive $B^\pm \to K^\pm \pi^+ \pi^- J/\psi$ decays}},}\
  }\href {\doibase 10.1103/PhysRevLett.91.262001} {\bibfield  {journal}
  {\bibinfo  {journal} {Phys. Rev. Lett.}\ }\textbf {\bibinfo {volume} {91}},\
  \bibinfo {pages} {262001} (\bibinfo {year} {2003})}\BibitemShut {NoStop}%
\bibitem [{\citenamefont {Aaboud}\ \emph {et~al.}(2017)\citenamefont {Aaboud}
  \emph {et~al.}}]{Aaboud:2016vzw}%
  \BibitemOpen
  \bibfield  {author} {\bibinfo {author} {\bibfnamefont {M.}~\bibnamefont
  {Aaboud}} \emph {et~al.} (\bibinfo {collaboration} {ATLAS}),\ }\bibfield
  {title} {\enquote {\bibinfo {title} {{Measurements of $\psi(2S)$ and $X(3872)
  \to J/\psi\pi^+\pi^-$ production in $pp$ collisions at $\sqrt{s} = 8$ TeV
  with the ATLAS detector}},}\ }\href {\doibase 10.1007/JHEP01(2017)117}
  {\bibfield  {journal} {\bibinfo  {journal} {J. High Energy Phys.}\ }\textbf {\bibinfo
  {volume} {01}},\ \bibinfo {pages} {117} (\bibinfo {year} {2017})}\BibitemShut
  {NoStop}%
\bibitem [{\citenamefont {Aubert}\ \emph {et~al.}(2005)\citenamefont {Aubert}
  \emph {et~al.}}]{Aubert:2004ns}%
  \BibitemOpen
  \bibfield  {author} {\bibinfo {author} {\bibfnamefont {B.}~\bibnamefont
  {Aubert}} \emph {et~al.} (\bibinfo {collaboration} {BaBar}),\ }\bibfield
  {title} {\enquote {\bibinfo {title} {{Study of the $B \to J/\psi K^- \pi^+
  \pi^-$ decay and measurement of the $B \to X(3872) K^-$ branching
  fraction}},}\ }\href {\doibase 10.1103/PhysRevD.71.071103} {\bibfield
  {journal} {\bibinfo  {journal} {Phys. Rev.}\ }\textbf {\bibinfo {volume}
  {D 71}},\ \bibinfo {pages} {071103} (\bibinfo {year} {2005})}\BibitemShut
  {NoStop}%
\bibitem [{\citenamefont {Ablikim}\ \emph {et~al.}(2014)\citenamefont {Ablikim}
  \emph {et~al.}}]{Ablikim:2013dyn}%
  \BibitemOpen
  \bibfield  {author} {\bibinfo {author} {\bibfnamefont {M.}~\bibnamefont
  {Ablikim}} \emph {et~al.} (\bibinfo {collaboration} {BESIII}),\ }\bibfield
  {title} {\enquote {\bibinfo {title} {{Observation of $e^+e^- \to \gamma
  X$(3872) at BES III}},}\ }\href {\doibase 10.1103/PhysRevLett.112.092001}
  {\bibfield  {journal} {\bibinfo  {journal} {Phys. Rev. Lett.}\ }\textbf
  {\bibinfo {volume} {112}},\ \bibinfo {pages} {092001} (\bibinfo {year}
  {2014})}\BibitemShut {NoStop}%
\bibitem [{\citenamefont {Acosta}\ \emph {et~al.}(2004)\citenamefont {Acosta}
  \emph {et~al.}}]{Acosta:2003zx}%
  \BibitemOpen
  \bibfield  {author} {\bibinfo {author} {\bibfnamefont {D.}~\bibnamefont
  {Acosta}} \emph {et~al.} (\bibinfo {collaboration} {CDF}),\ }\bibfield
  {title} {\enquote {\bibinfo {title} {{Observation of the narrow state
  $X(3872) \to J/\psi \pi^+ \pi^-$ in $\bar{p}p$ collisions at $\sqrt{s} =
  1.96$ TeV}},}\ }\href {\doibase 10.1103/PhysRevLett.93.072001} {\bibfield
  {journal} {\bibinfo  {journal} {Phys. Rev. Lett.}\ }\textbf {\bibinfo
  {volume} {93}},\ \bibinfo {pages} {072001} (\bibinfo {year}
  {2004})}\BibitemShut {NoStop}%
\bibitem [{\citenamefont {Chatrchyan}\ \emph {et~al.}(2013)\citenamefont
  {Chatrchyan} \emph {et~al.}}]{Chatrchyan:2013cld}%
  \BibitemOpen
  \bibfield  {author} {\bibinfo {author} {\bibfnamefont {S.}~\bibnamefont
  {Chatrchyan}} \emph {et~al.} (\bibinfo {collaboration} {CMS}),\ }\bibfield
  {title} {\enquote {\bibinfo {title} {{Measurement of the X(3872) production
  cross section via decays to $J/\psi \pi \pi$ in pp collisions at $\sqrt s$ =
  7 TeV}},}\ }\href {\doibase 10.1007/JHEP04(2013)154} {\bibfield  {journal}
  {\bibinfo  {journal} {J. High Energy Phys.}\ }\textbf {\bibinfo {volume} {04}},\ \bibinfo
  {pages} {154} (\bibinfo {year} {2013})}\BibitemShut {NoStop}%
\bibitem [{\citenamefont {Abazov}\ \emph {et~al.}(2004)\citenamefont {Abazov}
  \emph {et~al.}}]{Abazov:2004kp}%
  \BibitemOpen
  \bibfield  {author} {\bibinfo {author} {\bibfnamefont {V.~M.}\ \bibnamefont
  {Abazov}} \emph {et~al.} (\bibinfo {collaboration} {D0}),\ }\bibfield
  {title} {\enquote {\bibinfo {title} {{Observation and properties of the
  $X(3872)$ decaying to $J/\psi \pi^+ \pi^-$ in $p\bar{p}$ collisions at
  $\sqrt{s} = 1.96$ TeV}},}\ }\href {\doibase 10.1103/PhysRevLett.93.162002}
  {\bibfield  {journal} {\bibinfo  {journal} {Phys. Rev. Lett.}\ }\textbf
  {\bibinfo {volume} {93}},\ \bibinfo {pages} {162002} (\bibinfo {year}
  {2004})}\BibitemShut {NoStop}%
\bibitem [{\citenamefont {Aaij}\ \emph {et~al.}(2013)\citenamefont {Aaij} \emph
  {et~al.}}]{Aaij:2013zoa}%
  \BibitemOpen
  \bibfield  {author} {\bibinfo {author} {\bibfnamefont {R.}~\bibnamefont
  {Aaij}} \emph {et~al.} (\bibinfo {collaboration} {LHCb}),\ }\bibfield
  {title} {\enquote {\bibinfo {title} {{Determination of the X(3872) meson
  quantum numbers}},}\ }\href {\doibase 10.1103/PhysRevLett.110.222001}
  {\bibfield  {journal} {\bibinfo  {journal} {Phys. Rev. Lett.}\ }\textbf
  {\bibinfo {volume} {110}},\ \bibinfo {pages} {222001} (\bibinfo {year}
  {2013})}\BibitemShut {NoStop}%
\bibitem [{\citenamefont {Abazov}\ \emph {et~al.}(2016)\citenamefont {Abazov}
  \emph {et~al.}}]{D0:2016mwd}%
  \BibitemOpen
  \bibfield  {author} {\bibinfo {author} {\bibfnamefont {V.~M.}\ \bibnamefont
  {Abazov}} \emph {et~al.} (\bibinfo {collaboration} {D0}),\ }\bibfield
  {title} {\enquote {\bibinfo {title} {{Evidence for a $B_s^0 \pi^\pm$
  state}},}\ }\href {\doibase 10.1103/PhysRevLett.117.022003} {\bibfield
  {journal} {\bibinfo  {journal} {Phys. Rev. Lett.}\ }\textbf {\bibinfo
  {volume} {117}},\ \bibinfo {pages} {022003} (\bibinfo {year}
  {2016})}\BibitemShut {NoStop}%
\bibitem [{\citenamefont {Lyons}(2008)}]{lyons2008}%
  \BibitemOpen
  \bibfield  {author} {\bibinfo {author} {\bibfnamefont {L.}~\bibnamefont
  {Lyons}},\ }\bibfield  {title} {\enquote {\bibinfo {title} {Open statistical
  issues in particle physics},}\ }\href {\doibase 10.1214/08-AOAS163}
  {\bibfield  {journal} {\bibinfo  {journal} {Ann. Appl. Stat.}\ }\textbf
  {\bibinfo {volume} {2}},\ \bibinfo {pages} {887} (\bibinfo {year}
  {2008})}\BibitemShut {NoStop}%
\bibitem [{Note1()}]{Note1}%
  \BibitemOpen
  \bibinfo {note} {$\eta =-\protect \qopname \relax o{ln}[\protect \qopname
  \relax o{tan}(\theta /2)]\ $ is the pseudorapidity and $\theta $ is the polar
  angle between the track momentum and the proton beam direction. $\phi $ is
  the azimuthal angle of the track.}\BibitemShut {Stop}%
\bibitem [{\citenamefont {Aaij}\ \emph {et~al.}(2016)\citenamefont {Aaij} \emph
  {et~al.}}]{Aaij:2016iev}%
  \BibitemOpen
  \bibfield  {author} {\bibinfo {author} {\bibfnamefont {R.}~\bibnamefont
  {Aaij}} \emph {et~al.} (\bibinfo {collaboration} {LHCb}),\ }\bibfield
  {title} {\enquote {\bibinfo {title} {{Search for Structure in the
  $B_s^0\pi^\pm$ Invariant Mass Spectrum}},}\ }\href {\doibase
  10.1103/PhysRevLett.118.109904, 10.1103/PhysRevLett.117.152003} {\bibfield
  {journal} {\bibinfo  {journal} {Phys. Rev. Lett.}\ }\textbf {\bibinfo
  {volume} {117}},\ \bibinfo {pages} {152003} (\bibinfo {year} {2016})},\
  \bibinfo {note} {[Addendum: Phys. Rev. Lett. {\bf 118}, 109904
  (2017)]}\BibitemShut {NoStop}%
\bibitem [{\citenamefont {Sirunyan}\ \emph {et~al.}(2017)\citenamefont
  {Sirunyan} \emph {et~al.}}]{Sirunyan:2017ofq}%
  \BibitemOpen
  \bibfield  {author} {\bibinfo {author} {\bibfnamefont {A.~M.}\ \bibnamefont
  {Sirunyan}} \emph {et~al.} (\bibinfo {collaboration} {CMS}),\ }\bibfield
  {title} {\enquote {\bibinfo {title} {{Search for the X(5568) state decaying
  into $\mathrm{B}^{0}_{\mathrm{s}}\pi^{\pm}$ in proton-proton collisions at
  $\sqrt{s} = $ 8 TeV}},}\ }\href@noop {} {\  (\bibinfo {year} {2017})},\
  \Eprint {http://arxiv.org/abs/1712.06144} {arXiv:1712.06144}
  \BibitemShut {NoStop}%
\bibitem [{\citenamefont {Aaltonen}\ \emph {et~al.}(2017)\citenamefont
  {Aaltonen} \emph {et~al.}}]{Aaltonen:2017voc}%
  \BibitemOpen
  \bibfield  {author} {\bibinfo {author} {\bibfnamefont {T.}~\bibnamefont
  {Aaltonen}} \emph {et~al.} (\bibinfo {collaboration} {CDF}),\ }\bibfield
  {title} {\enquote {\bibinfo {title} {{A search for the exotic meson $X(5568)$
  with the Collider Detector at Fermilab}},}\ }\href@noop {} {\  (\bibinfo
  {year} {2017})},\ \Eprint {http://arxiv.org/abs/1712.09620} {arXiv:1712.09620
  [hep-ex]} \BibitemShut {NoStop}%
\bibitem [{\citenamefont {Abazov}\ \emph {et~al.}(2006)\citenamefont {Abazov}
  \emph {et~al.}}]{Abazov:2005pn}%
  \BibitemOpen
  \bibfield  {author} {\bibinfo {author} {\bibfnamefont {V.~M.}\ \bibnamefont
  {Abazov}} \emph {et~al.} (\bibinfo {collaboration} {D0}),\ }\bibfield
  {title} {\enquote {\bibinfo {title} {{The Upgraded D0 detector}},}\ }\href
  {\doibase 10.1016/j.nima.2006.05.248} {\bibfield  {journal} {\bibinfo
  {journal} {Nucl. Instrum. Meth.}\ }\textbf {\bibinfo {volume} {A565}},\
  \bibinfo {pages} {463--537} (\bibinfo {year} {2006})}\BibitemShut {NoStop}%
\bibitem [{\citenamefont {Angstadt}\ \emph {et~al.}(2010)\citenamefont
  {Angstadt} \emph {et~al.}}]{Angstadt:2009ie}%
  \BibitemOpen
  \bibfield  {author} {\bibinfo {author} {\bibfnamefont {R.}~\bibnamefont
  {Angstadt}} \emph {et~al.} (\bibinfo {collaboration} {D0}),\ }\bibfield
  {title} {\enquote {\bibinfo {title} {{The Layer 0 Inner Silicon Detector of
  the D0 Experiment}},}\ }\href {\doibase 10.1016/j.nima.2010.04.148}
  {\bibfield  {journal} {\bibinfo  {journal} {Nucl. Instrum. Meth.}\ }\textbf
  {\bibinfo {volume} {A622}},\ \bibinfo {pages} {298--310} (\bibinfo {year}
  {2010})}\BibitemShut {NoStop}%
\bibitem [{\citenamefont {Abazov}\ \emph {et~al.}(2005)\citenamefont {Abazov}
  \emph {et~al.}}]{Abazov:2005uk}%
  \BibitemOpen
  \bibfield  {author} {\bibinfo {author} {\bibfnamefont {V.~M.}\ \bibnamefont
  {Abazov}} \emph {et~al.},\ }\bibfield  {title} {\enquote {\bibinfo {title}
  {{The Muon system of the Run II D0 detector}},}\ }\href {\doibase
  10.1016/j.nima.2005.07.008} {\bibfield  {journal} {\bibinfo  {journal} {Nucl.
  Instrum. Meth.}\ }\textbf {\bibinfo {volume} {A552}},\ \bibinfo {pages}
  {372--398} (\bibinfo {year} {2005})}\BibitemShut {NoStop}%
\bibitem [{\citenamefont {Abazov}\ \emph {et~al.}(2013)\citenamefont {Abazov}
  \emph {et~al.}}]{Abazov:2012zz}%
  \BibitemOpen
  \bibfield  {author} {\bibinfo {author} {\bibfnamefont {V.~M.}\ \bibnamefont
  {Abazov}} \emph {et~al.} (\bibinfo {collaboration} {D0}),\ }\bibfield
  {title} {\enquote {\bibinfo {title} {{Measurement of the Semileptonic Charge
  Asymmetry using $B_s^0 \to D_s\mu X$ Decays}},}\ }\href {\doibase
  10.1103/PhysRevLett.110.011801} {\bibfield  {journal} {\bibinfo  {journal}
  {Phys. Rev. Lett.}\ }\textbf {\bibinfo {volume} {110}},\ \bibinfo {pages}
  {011801} (\bibinfo {year} {2013})}\BibitemShut {NoStop}%
\bibitem [{\citenamefont {Abdallah}\ \emph {et~al.}(2004)\citenamefont
  {Abdallah} \emph {et~al.}}]{Abdallah:2002xm}%
  \BibitemOpen
  \bibfield  {author} {\bibinfo {author} {\bibfnamefont {J.}~\bibnamefont
  {Abdallah}} \emph {et~al.} (\bibinfo {collaboration} {DELPHI}),\ }\bibfield
  {title} {\enquote {\bibinfo {title} {{b tagging in DELPHI at LEP}},}\ }\href
  {\doibase 10.1140/epjc/s2003-01441-8} {\bibfield  {journal} {\bibinfo
  {journal} {Eur. Phys. J.}\ }\textbf {\bibinfo {volume} {C32}},\ \bibinfo
  {pages} {185--208} (\bibinfo {year} {2004})}\BibitemShut {NoStop}%
\bibitem [{Note2()}]{Note2}%
  \BibitemOpen
  \bibinfo {note} {The three dimensional impact parameter (IP) is defined as
  the distance of closest approach of the track to the $p\protect \mathaccentV
  {bar}016{p}$ collision point. The two dimensional IP is the distance of
  closest approach projected onto the plane transverse to the $p\protect
  \mathaccentV {bar}016{p}$ beams.}\BibitemShut {Stop}%
\bibitem [{\citenamefont {Patrignani}\ \emph {et~al.}(2016)\citenamefont
  {Patrignani} \emph {et~al.}}]{Olive:2016xmw}%
  \BibitemOpen
  \bibfield  {author} {\bibinfo {author} {\bibfnamefont {C.}~\bibnamefont
  {Patrignani}} \emph {et~al.} (\bibinfo {collaboration} {Particle Data
  Group}),\ }\bibfield  {title} {\enquote {\bibinfo {title} {{Review of
  Particle Physics}},}\ }\href {\doibase 10.1088/1674-1137/40/10/100001}
  {\bibfield  {journal} {\bibinfo  {journal} {Chin. Phys.}\ }\textbf {\bibinfo
  {volume} {C40}},\ \bibinfo {pages} {100001} (\bibinfo {year}
  {2016})}\BibitemShut {NoStop}%
\bibitem [{\citenamefont {Sjostrand}\ \emph {et~al.}(2006)\citenamefont
  {Sjostrand}, \citenamefont {Mrenna},\ and\ \citenamefont
  {Skands}}]{Sjostrand:2006za}%
  \BibitemOpen
  \bibfield  {author} {\bibinfo {author} {\bibfnamefont {T.}~\bibnamefont
  {Sjostrand}}, \bibinfo {author} {\bibfnamefont {S.}~\bibnamefont {Mrenna}}, \
  and\ \bibinfo {author} {\bibfnamefont {P.Z.}\ \bibnamefont {Skands}},\
  }\bibfield  {title} {\enquote {\bibinfo {title} {{PYTHIA 6.4 Physics and
  Manual}},}\ }\href {\doibase 10.1088/1126-6708/2006/05/026} {\bibfield
  {journal} {\bibinfo  {journal} {JHEP}\ }\textbf {\bibinfo {volume} {05}},\
  \bibinfo {pages} {026} (\bibinfo {year} {2006})}\BibitemShut {NoStop}%
\bibitem [{\citenamefont {Lange}(2001)}]{Lange:2001uf}%
  \BibitemOpen
  \bibfield  {author} {\bibinfo {author} {\bibfnamefont {D.~J.}\ \bibnamefont
  {Lange}},\ }\bibfield  {title} {\enquote {\bibinfo {title} {{The EvtGen
  particle decay simulation package}},}\ }\bibfield  {booktitle} {\emph
  {\bibinfo {booktitle} {{Proceedings, 7th International Conference on B
  physics at hadron machines (BEAUTY 2000): Maagan, Israel, September 13-18,
  2000}}},\ }\href {\doibase 10.1016/S0168-9002(01)00089-4} {\bibfield
  {journal} {\bibinfo  {journal} {Nucl. Instrum. Meth.}\ }\textbf {\bibinfo
  {volume} {A462}},\ \bibinfo {pages} {152--155} (\bibinfo {year}
  {2001})}\BibitemShut {NoStop}%
\bibitem [{Pro(1974)}]{Proceedings:1974sfa}%
  \BibitemOpen
  \href {\doibase 10.5170/CERN-1974-023} {\emph {\bibinfo {title} {{1974 CERN
  School of Computing, Godoysund, Norway, 11-24 Aug 1974: Proceedings}}}}\
  (\bibinfo {year} {1974})\BibitemShut {NoStop}%
\bibitem [{\citenamefont {Albrecht}\ \emph {et~al.}(1990)\citenamefont
  {Albrecht} \emph {et~al.}}]{Albrecht:1990am}%
  \BibitemOpen
  \bibfield  {author} {\bibinfo {author} {\bibfnamefont {H.}~\bibnamefont
  {Albrecht}} \emph {et~al.} (\bibinfo {collaboration} {ARGUS}),\ }\bibfield
  {title} {\enquote {\bibinfo {title} {{Search for Hadronic $b \to u$
  Decays}},}\ }\href {\doibase 10.1016/0370-2693(90)91293-K} {\bibfield
  {journal} {\bibinfo  {journal} {Phys. Lett.}\ }\textbf {\bibinfo {volume}
  {B241}},\ \bibinfo {pages} {278--282} (\bibinfo {year} {1990})}\BibitemShut
  {NoStop}%
\bibitem [{\citenamefont {Giunti}(1999)}]{Giunti:1998xv}%
  \BibitemOpen
  \bibfield  {author} {\bibinfo {author} {\bibfnamefont {C.}~\bibnamefont
  {Giunti}},\ }\bibfield  {title} {\enquote {\bibinfo {title} {{Treatment of
  the background error in the statistical analysis of Poisson processes}},}\
  }\href {\doibase 10.1103/PhysRevD.59.113009} {\bibfield  {journal} {\bibinfo
  {journal} {Phys. Rev.}\ }\textbf {\bibinfo {volume} {D59}},\ \bibinfo {pages}
  {113009} (\bibinfo {year} {1999})}\BibitemShut {NoStop}%
\bibitem [{\citenamefont {Gross}\ and\ \citenamefont
  {Vitells}(2010)}]{Gross:2010qma}%
  \BibitemOpen
  \bibfield  {author} {\bibinfo {author} {\bibfnamefont {E.}~\bibnamefont
  {Gross}}\ and\ \bibinfo {author} {\bibfnamefont {O.}~\bibnamefont
  {Vitells}},\ }\bibfield  {title} {\enquote {\bibinfo {title} {{Trial factors
  or the look elsewhere effect in high energy physics}},}\ }\href {\doibase
  10.1140/epjc/s10052-010-1470-8} {\bibfield  {journal} {\bibinfo  {journal}
  {Eur. Phys. J.}\ }\textbf {\bibinfo {volume} {C70}},\ \bibinfo {pages}
  {525--530} (\bibinfo {year} {2010})}\BibitemShut {NoStop}%
\end{thebibliography}

%

\end{document}